\documentclass[twocolumn,tighten,trackchanges]{aastex631}

\shorttitle{Magnetospheric current sheet alignment}
\shortauthors{Selvi et al.}

\graphicspath{{./}{figures/}}

\usepackage{longtable}

\usepackage{array,booktabs}
\usepackage[symbols,nogroupskip,nonumberlist]{glossaries-extra}

\usepackage[normalem]{ulem}
\usepackage{soul}
\usepackage{array}
\usepackage{xcolor}

\glsdisablehyper
\glsxtrnewsymbol[description={Speed of light}]{spol}{\ensuremath{c}}
\glsxtrnewsymbol[description={Time}]{time}{\ensuremath{t}}
\glsxtrnewsymbol[description={Number density}]{ndens}{\ensuremath{n}}
\glsxtrnewsymbol[description={Number}]{num}{\ensuremath{N}}
\glsxtrnewsymbol[description={Volume}]{vol}{\ensuremath{V}}
\glsxtrnewsymbol[description={Mass}]{mass}{\ensuremath{M}}
\glsxtrnewsymbol[description={Mass density}]{mdens}{\ensuremath{\rho}}
\glsxtrnewsymbol[description={Generalized coordinates}]{gcoords}{\ensuremath{\textbf{q}}}
\glsxtrnewsymbol[description={Generalized coordinate}]{gcoord}{\ensuremath{q}}
\glsxtrnewsymbol[description={Magnetic field energy density}]{magnenrgdens}{\ensuremath{\mathcal{E}_{\mathrm{B}}  }}
\glsxtrnewsymbol[description={Magnetic field components}]{Bcomp}{\ensuremath{B}}
\glsxtrnewsymbol[description={Magnetic field 3-vector}]{B}{\ensuremath{\textbf{B}}}
\glsxtrnewsymbol[description={Electric field energy density}]{elecenrgdens}{\ensuremath{\mathcal{E}_{\mathrm{E}}}}
\glsxtrnewsymbol[description={Electric field components}]{Ecomp}{\ensuremath{E}}
\glsxtrnewsymbol[description={Electric field 3-vector}]{E}{\ensuremath{\textbf{E}}}
\glsxtrnewsymbol[description={Electromagnetic field energy density}]{elecmagnenrgdens}{\ensuremath{\mathcal{E}_{\mathrm{EM}}}}
\glsxtrnewsymbol[description={Lorentz factor}]{lfacp}{\ensuremath{\gamma}}
\glsxtrnewsymbol[description={Lorentz factor}]{lfacbulk}{\ensuremath{\Gamma}}
\glsxtrnewsymbol[description={Particle mass}]{pmass}{\ensuremath{m_{p}}}
\glsxtrnewsymbol[description={Particle energy}]{prtenrgdens}{\ensuremath{E_{\mathrm{p}}}}
\glsxtrnewsymbol[description={Particle rest-mass energy}]{rmenrgdens}{\ensuremath{E_{\mathrm{rm}}}}
\glsxtrnewsymbol[description={Particle kinetic energy}]{kinenrgdens}{\ensuremath{E_{\mathrm{k}}}}
\glsxtrnewsymbol[description={3-velocity components}]{3velcomp}{\ensuremath{v}}
\glsxtrnewsymbol[description={3-velocity}]{3vel}{\ensuremath{\textbf{v}}}
\glsxtrnewsymbol[description={4-velocity components}]{4velcomp}{\ensuremath{u}}
\glsxtrnewsymbol[description={4-momentum components}]{4momcomp}{\ensuremath{p}}
\glsxtrnewsymbol[description={4-velocity}]{4vel}{\ensuremath{\textbf{u}}}
\glsxtrnewsymbol[description={4-momentum}]{4mom}{\ensuremath{\textbf{p}}}
\glsxtrnewsymbol[description={Energy-momentum tensor}]{enrgmomten}{\ensuremath{\textbf{T}}}
\glsxtrnewsymbol[description={Energy-momentum tensor components}]{enrgmomtencomp}{\ensuremath{T}}
\glsxtrnewsymbol[description={Energy-momentum tensor density components}]{enrgmomtendenscomp}{\ensuremath{\mathfrak{T}}}
\glsxtrnewsymbol[description={4-position}]{4pos}{\ensuremath{\mathbf{x}}}
\glsxtrnewsymbol[description={4-position components}]{4poscomp}{\ensuremath{x}}
\glsxtrnewsymbol[description={Metric tensor}]{metr}{\ensuremath{\textbf{g}}}
\glsxtrnewsymbol[description={Metric tensor components}]{metrcomp}{\ensuremath{g}}
\glsxtrnewsymbol[description={Affine parameter}]{affpar}{\ensuremath{s}}
\glsxtrnewsymbol[description={System distribution function}]{distrfunc}{\ensuremath{f}}
\glsxtrnewsymbol[description={Metric tensor}]{metrten}{\ensuremath{\textbf{g}}}
\glsxtrnewsymbol[description={Metric tensor components}]{metrtencomp}{\ensuremath{\tilde{g}}}
\glsxtrnewsymbol[description={Metric tensor determinant}]{metrtendet}{\ensuremath{g}}
\glsxtrnewsymbol[description={Minkowksi metric tensor components}]{minkmetrtencomp}{\ensuremath{\eta}}
\glsxtrnewsymbol[description={Cartesian x coordinate}]{cartx}{\ensuremath{x}}
\glsxtrnewsymbol[description={Cartesian y coordinate}]{carty}{\ensuremath{y}}
\glsxtrnewsymbol[description={Cartesian z coordinate}]{cartz}{\ensuremath{z}}
\glsxtrnewsymbol[description={Spherical radial coordinate}]{sphradius}{\ensuremath{r}}
\glsxtrnewsymbol[description={Spherical polar coordinate}]{sphpolar}{\ensuremath{\theta}}
\glsxtrnewsymbol[description={Spherical azimuthal coordinate}]{sphazi}{\ensuremath{\phi}}
\glsxtrnewsymbol[description={Schwarzschild radius}]{schwradius}{\ensuremath{R_{s}}}
\glsxtrnewsymbol[description={Graviational constant}]{gravconst}{\ensuremath{G}}
\glsxtrnewsymbol[description={Black hole spin parameter}]{spinpar}{\ensuremath{a}}
\glsxtrnewsymbol[description={Magnetic vector potential}]{magnvecpot}{\ensuremath{\textbf{A}}}
\glsxtrnewsymbol[description={Magnetic vector potential components}]{magnvecpotcomp}{\ensuremath{A}}
\glsxtrnewsymbol[description={Current}]{curr}{\ensuremath{I}}
\glsxtrnewsymbol[description={Vacuum mangetic permeability}]{vacmagnperm}{\ensuremath{\mu_{0}}}
\glsxtrnewsymbol[description={Magnetic dipole moment}]{magndipmom}{\ensuremath{ \textbf{m} }}

\glsxtrnewsymbol[description={Electromagnetic 4-potential}]{4pot}{\ensuremath{\textbf{A}}} 
\glsxtrnewsymbol[description={Electromagnetic 4-potential components}]{4potcomp}{\ensuremath{A}} 
\glsxtrnewsymbol[description={Electric scalar potential}]{elecpot}{\ensuremath{ \tilde{\phi}  }} 
\glsxtrnewsymbol[description={Electromagnetic field tensor}]{emtens}{\ensuremath{\textbf{F}}} 
\glsxtrnewsymbol[description={Electromagnetic field tensor components}]{emtenscomp}{\ensuremath{F}} 
\glsxtrnewsymbol[description={Electrical resistivity}]{elres}{\ensuremath{\eta}} 
\glsxtrnewsymbol[description={Magnetic diffusivity}]{mdiff}{\ensuremath{\tilde{\eta}}} 
\glsxtrnewsymbol[description={Electrical conductance}]{elcond}{\ensuremath{\sigma_{e}}} 
\glsxtrnewsymbol[description={Modified Kerr-Schild radius stretching parameter}]{mksr}{\ensuremath{R_{0,MKS}}} 
\glsxtrnewsymbol[description={Modified Kerr-Schild polar stretching parameter}]{mksh}{\ensuremath{h_{0,MKS}}} 
\glsxtrnewsymbol[description={Plasma magnetization}]{plmagn}{\ensuremath{\sigma}} 
\glsxtrnewsymbol[description={Electromagnetic vector potential}]{vecpot}{\ensuremath{\textrm{A} }} 
\glsxtrnewsymbol[description={Electromagnetic vector potential components}]{vecpotcomp}{\ensuremath{A}} 
\glsxtrnewsymbol[description={Magnetic moment}]{magnmom}{\ensuremath{\mu}} 
\glsxtrnewsymbol[description={Pressure}]{pres}{\ensuremath{p}} 
\glsxtrnewsymbol[description={Temperature}]{temp}{\ensuremath{T}} 

\glsxtrnewsymbol[description={Magnetopheric tilt angle}]{tiltangle}{\ensuremath{\chi}} 
\glsxtrnewsymbol[description={Length}]{len}{\ensuremath{\mathcal{L}}} 
\glsxtrnewsymbol[description={Resolution}]{res}{\ensuremath{\mathcal{R}}} 
\glsxtrnewsymbol[description={Plasma beta}]{plbeta}{\ensuremath{\beta}} 

\glsxtrnewsymbol[description={Fluid pressure}]{flpr}{\ensuremath{p_{f}}} 
\glsxtrnewsymbol[description={Polytropic index}]{polyidx}{\ensuremath{\hat{\gamma}}} 
\glsxtrnewsymbol[description={Entropy related quantity}]{entralt}{\ensuremath{S}} 
\glsxtrnewsymbol[description={Entropy related quantity}]{boltzcon}{\ensuremath{k_{\mathrm{B}}}} 
\glsxtrnewsymbol[description={Dimensionless temperature/thermal spread}]{dimtemp}{\ensuremath{\Theta}} 
\glsxtrnewsymbol[description={Magnetic pressure}]{magnpr}{\ensuremath{p_{\mathrm{m}}}} 
\glsxtrnewsymbol[description={Vector potential constant}]{vecpotcon}{\ensuremath{g}} 
\glsxtrnewsymbol[description={Magnetic inclination angle}]{magninclangle}{\ensuremath{\chi}} 
\glsxtrnewsymbol[description={Relativistic enthaply density}]{relenthdens}{\ensuremath{w}} 
\glsxtrnewsymbol[description={Gravitational radius}]{gravradius}{\ensuremath{r_{\textrm{g}}  }} 
\glsxtrnewsymbol[description={Angular velocity}]{angfreq}{\ensuremath{ \Omega}} 
\glsxtrnewsymbol[description={Volume 3-current density}]{3currdenscomp}{\ensuremath{J}}
\glsxtrnewsymbol[description={Magnetic flux}]{magnflux}{\ensuremath{ \Phi_{\textrm{B}}   }}
\glsxtrnewsymbol[description={Volume 3-current density}]{3currdens}{\ensuremath{\textbf{J}}}
\glsxtrnewsymbol[description={Alfv{\'e}n speed}]{alfvensp}{\ensuremath{  v_{\textrm{A}}   }}
\glsxtrnewsymbol[description={Drift velocity}]{driftvel}{\ensuremath{  v_{\textrm{d}}   }}
\glsxtrnewsymbol[description={Spatial metric determinant}]{spmetricdet}{\ensuremath{  \gamma  }}
\glsxtrnewsymbol[description={Radial MKS coordinate}]{mksradius}{\ensuremath{  s  }}
\glsxtrnewsymbol[description={Lapse function}]{lapse}{\ensuremath{  \alpha  }}
\glsxtrnewsymbol[description={Power}]{power}{\ensuremath{  P  }}
\glsxtrnewsymbol[description={Energy flux}]{enrgyflux}{\ensuremath{  \dot{E} }}
\glsxtrnewsymbol[description={Event horizon radius}]{ehradius}{\ensuremath{   r_{\textrm{H} } }} 
\glsxtrnewsymbol[description={Magnetic flux}]{magnfluxabs}{\ensuremath{ \Phi_{|\textrm{B}|}   }}
\glsxtrnewsymbol[description={Angular momentum}]{angmom}{\ensuremath{\textbf{L}}} 
\glsxtrnewsymbol[description={Comoving magnetic field components}]{bcomp}{\ensuremath{b}}
\glsxtrnewsymbol[description={Comoving magnetic field 3-vector}]{b}{\ensuremath{\textbf{b}}}
\glsxtrnewsymbol[description={Period}]{period}{\ensuremath{P}}
\glsxtrnewsymbol[description={Elementary charge}]{elch}{\ensuremath{e}}
\glsxtrnewsymbol[description={Mulitplicity}]{mplcty}{\ensuremath{\lambda}}
\glsxtrnewsymbol[description={Force}]{force}{\ensuremath{F}}
\glsxtrnewsymbol[description={Reconnection rate}]{recrate}{\ensuremath{R}}
\glsxtrnewsymbol[description={Energy density}]{enrgdens}{\ensuremath{U}}
\glsxtrnewsymbol[description={Thomson cross section}]{csthomson}{\ensuremath{\sigma_{\textrm{T}}}}
\glsxtrnewsymbol[description={Length}]{length}{\ensuremath{L}}
\glsxtrnewsymbol[description={Energy}]{enrg}{\ensuremath{\mathcal{E}}}
\glsxtrnewsymbol[description={Relativistic specific enthaply}]{relspecenth}{\ensuremath{h}}

\newcommand*{\rom}[1]{\expandafter\@slowromancap\romannumeral #1@}

\begin{document}

%%%%%%%%%%%%%%%%%%%%%%%%%%%%%%%%%%%%%%%%%%%%%%%
%%%                    Title                %%%
%%%%%%%%%%%%%%%%%%%%%%%%%%%%%%%%%%%%%%%%%%%%%%%
 
\title{Current sheet alignment in oblique black hole magnetospheres -- a black hole pulsar?}

%%%%%%%%%%%%%%%%%%%%%%%%%%%%%%%%%%%%%%%%%%%%%%%
%%%                   Authors               %%%
%%%%%%%%%%%%%%%%%%%%%%%%%%%%%%%%%%%%%%%%%%%%%%%
\author[0000-0001-9508-1234]{S. Selvi}
\affiliation{Anton Pannekoek Institute, Science Park 904, 1098 XH, Amsterdam, The Netherlands}

\author[0000-0002-4584-2557]{O. Porth}
\affiliation{Anton Pannekoek Institute, Science Park 904, 1098 XH, Amsterdam, The Netherlands}

\author[0000-0002-7301-3908]{B. Ripperda}

\affiliation{Canadian Institute for Theoretical Astrophysics, 60 St. George Street, Toronto, Ontario M5S 3H8}
\affiliation{Department of Physics, University of Toronto, 60 St. George Street, Toronto, ON M5S 1A7}
\affiliation{David A. Dunlap Department of Astronomy, University of Toronto, 50 St. George Street, Toronto, ON M5S 3H4}
\affiliation{Perimeter Institute for Theoretical Physics, 31 Caroline St. North, Waterloo, ON, Canada N2L 2Y5}
\affiliation{Center for Computational Astrophysics, Flatiron Institute, 162 5th avenue, New York, NY, 10010, USA}

\author[0000-0002-1227-2754]{L. Sironi}

\affiliation{Department of Astronomy, Columbia University, 550 West 120th street, New York, NY, 10027, USA}
\affiliation{Center for Computational Astrophysics, Flatiron Institute, 162 5th avenue, New York, NY, 10010, USA}

% \author{A. Philippov}
% \affiliation{affiliation}

%%%%%%%%%%%%%%%%%%%%%%%%%%%%%%%%%%%%%%%%%%%%%%%
%%%                Abstract                 %%%
%%%%%%%%%%%%%%%%%%%%%%%%%%%%%%%%%%%%%%%%%%%%%%%
\begin{abstract}
We study the magnetospheric evolution of a non-accreting spinning black hole (BH) with an initially inclined split monopole magnetic field by means of three-dimensional general relativistic magnetohydrodynamics simulations. 
This serves as a model for a neutron star (NS) collapse or a BH-NS merger remnant after the inherited magnetosphere has settled into a split monopole field creating a striped wind.
We show that the initially inclined split monopolar current sheet aligns over time with the BH equatorial plane. 
The inclination angle evolves exponentially towards alignment, with an alignment timescale that is inversely proportional to the square of the BH angular velocity, where higher spin results in faster alignment.  
Furthermore, magnetic reconnection in the current sheet leads to exponential decay of event horizon penetrating magnetic flux with nearly the same timescale for all considered BH spins.
In addition, we present relations for the BH mass and spin in terms of the period and alignment timescale of the striped wind.
The explored scenario of a rotating, aligning and reconnecting current sheet can potentially lead to multimessenger electromagnetic counterparts to a gravitational wave event due to the acceleration of particles powering high-energy radiation, plasmoid mergers resulting in coherent radio signals, and pulsating emission due to the initial misalignment of the BH magnetosphere.
\end{abstract}

%%%%%%%%%%%%%%%%%%%%%%%%%%%%%%%%%%%%%%%%%%%%%%%
%%%                Keywords                 %%%
%%%%%%%%%%%%%%%%%%%%%%%%%%%%%%%%%%%%%%%%%%%%%%%
\keywords{
\href{http://astrothesaurus.org/uat/739}{High energy astrophysics (739)}; 
\href{http://astrothesaurus.org/uat/1261}{Plasma astrophysics (1261)}; 
\href{http://astrothesaurus.org/uat/288}{Compact objects (288)}; 
\href{http://astrothesaurus.org/uat/994}{Magnetic fields (994)}; 
\href{http://astrothesaurus.org/uat/1393}{Relativity (1393)}; 
\href{http://astrothesaurus.org/uat/1964}{Magnetohydrodynamics (1964)} 
}

%%%%%%%%%%%%%%%%%%%%%%%%%%%%%%%%%%%%%%%%%%%%%%%
%%%          Section Introduction           %%%
%%%%%%%%%%%%%%%%%%%%%%%%%%%%%%%%%%%%%%%%%%%%%%%
\section{Introduction} 
A black hole (BH) with a magnetic field penetrating its event horizon can be born from magnetized progenitors such as a spinning down rotationally supported hypermassive neutron star (NS) \citep{Falcke2014} or a BH-NS merger remnant \citep{East2021}.

According to the no-hair theorem \citep{Misner1973} the magnetic flux on a BH event horizon in vacuum will inevitably decay over time \citep{Price1972}.
However, BHs born from magnetized progenitors will, very likely, not be isolated in a vacuum but be surrounded by a (conductive) plasma, thereby changing the magnetospheric dynamics. 
In a stationary axisymmetric magnetosphere, poloidal magnetic field loops with their ends penetrating the event horizon can only exist if supported by non-force-free currents \citep{MacDonald1982,Gralla2014}. 
It follows that in a force-free magnetosphere, an inherited closed magnetic field geometry, e.g., typically assumed to be a dipole, will inevitably evolve into a split monopole geometry \citep{Komissarov2004,Lyutikov2011,Bransgrove2021} in which the current sheet extends all the way to the event horizon.
In the limit of infinite conductivity the magnetic field would be prevented from decaying. 
However in nature, although typically extremely high, a finite conductivity allows for magnetic reconnection leading to magnetic field decay \citep{Lyutikov2011,Bransgrove2021}.
This is in contrast to the case of a neutron star (NS), where reconnection occurs past the light-cylinder and the flux through the assumed perfectly conducting NS surface remains unchanged \citep{Goldreich1969,Bogovalov1999,Philippov2019}.

A strikingly similar situation of a BH shedding its event horizon magnetic flux through magnetic reconnection could occur in accreting BHs in the magnetically arrested disk (MAD) state and may result in observable powerful X-ray and gamma-ray flares \citep{Ripperda2020,Chashkina2021,Ripperda2022,Hakobyan2022}.

In previous studies BH magnetic and rotational axes were aligned, both in 2D \citep{Lyutikov2011} and in 3D \citep{Bransgrove2021}. 
These studies showed the formation of a split monopole geometry after an initial transient phase starting from a dipolar field.
Therefore, in this work we assume that a split monopole geometry has already formed on the event horizon.
Unlike what was assumed in earlier works, it is likely that the inherited magnetosphere is misaligned with the BH spin axis.
In order to study this more general scenario, we here consider a non-accreting spinning BH magnetosphere with an initially inclined split monopole geometry in 3D.

%%%%%%%%%%%%%%%%%%%%%%%%%%%%%%%%%%%%%%%%%%%%%%%
%%% Section Numerical methods and setup     %%%
%%%%%%%%%%%%%%%%%%%%%%%%%%%%%%%%%%%%%%%%%%%%%%%
\section{Numerical methods and setup} 
\label{sec:numericalmethods}
We adopt a general relativistic (ideal) magnetohydrodynamic (GRMHD) description of the magnetospheric plasma and employ a (fixed) Kerr spacetime geometry with a range of dimensionless BH spin parameters $\gls{spinpar} \in \{0.5,0.8,0.9,0.99\}$, BH mass $\gls{mass}$, gravitational constant $\gls{gravconst}$, speed of light $\gls{spol}$, and gravitational radius $\gls{gravradius} = \gls{gravconst} \gls{mass} / \gls{spol}^{2}$.
The simulations are performed using the BlackHoleAccretionCode (\texttt{BHAC})  \citep{Porth2017,Olivares2019}.
The GRMHD equations are solved in spherical, event horizon penetrating logarithmic Kerr-Schild coordinates  $(\gls{time}, \gls{mksradius},\gls{sphpolar},\gls{sphazi})$ \citep{McKinney2004,Porth2017} with $\gls{mksradius}=\textrm{ln}(\gls{sphradius})$, where $\gls{time}$ is time and $\gls{sphradius}$, $\gls{sphpolar}$, $\gls{sphazi}$ are the spatial spherical (Kerr-Schild) coordinates and the BH spin axis is aligned with the $\gls{cartz}$-axis (i.e. along $\gls{sphpolar} = 0$).
The Eulerian magnetic field components $\gls{Bcomp}^{i}$ (with i $\in {1,2,3}$) are initialized by a split monopole magnetic vector potential $\gls{vecpotcomp}_{\gls{sphazi}} =  \left(1-\sqrt{\textrm{cos}^{2} \gls{sphpolar}  }\right) / (\gls{sphradius}\, \textrm{sin}  \gls{sphpolar}  )$ rotated by an initial inclination angle $\chi_{0}  = 30^\circ $ about the $\gls{carty}$-axis.

We impose an equation of state for a fully relativistic ideal gas with an adiabatic index $\gls{polyidx} = 4/3$ initializing pressure as $\gls{pres}(\gls{sphradius}) = \gls{dimtemp} \gls{mdens}(\gls{sphradius}) \gls{spol}^{2}$ with a dimensionless temperature  $ \gls{dimtemp} = 0.05$ for a cold plasma and $\gls{mdens}$ the plasma density.

In order to closely represent a split monopole magnetosphere with $\gls{Bcomp} \sim \gls{sphradius}^{-2}$ and $\gls{mdens}(\gls{sphradius})  \sim \gls{sphradius}^{-2}$ the initial plasma magnetization is set to 
$
\gls{plmagn}(\gls{sphradius}) 
=
\gls{plmagn}_{0} \gls{sphradius}^{-2} 
=
\gls{Bcomp}^{2} /( \gls{mdens}(\gls{sphradius})  \gls{relspecenth})  
$ 
for $\gls{sphradius} \leq 2 \gls{gravradius}$ (i.e. inside a radius of the maximum extent of the ergoregion)
with $\gls{plmagn}_{0} = 50 $ and 
$\gls{relspecenth}
=
( 1 + 
 (\gls{polyidx}/(\gls{polyidx} -1)) 
\gls{dimtemp}  )   \gls{spol}^{2} 
$ the relativistic specific enthalpy.
To avoid the plasma magnetization from becoming so small for large radii that it would no longer represent a force-free magnetosphere we set the initial plasma magnetization for $\gls{sphradius} \geq 2 \gls{gravradius}$ to be uniform and equal to its value at $\gls{sphradius} = 2 \gls{gravradius}$ to
$\gls{plmagn} 
=
\gls{Bcomp}^{2} /( \gls{mdens} \gls{relspecenth})
=
12.5$. 
This plasma magnetization sets the initial plasma density field $\gls{mdens}(\gls{sphradius})$ and the initial plasma beta (i.e. gas-to-magnetic pressure ratio) 
$\gls{plbeta} (\gls{sphradius})  
=
2 \gls{dimtemp} / ( 1.2 \gls{plmagn} (\gls{sphradius}))
$ (i.e. $ \gls{plbeta} (\gls{sphradius} \geq 2 \gls{gravradius}) = 6.7 \cdot 10^{-3}$  
).

The computational domain covers $ 1.01  \leq \gls{sphradius} /  \gls{gravradius} \leq 400 $, $0 \leq \gls{sphpolar}  \leq \pi $, and $0 \leq \gls{sphazi}  < 2 \pi$ and we run all simulations until $\gls{time} = 400 \ \gls{gravradius} / \gls{spol}$.  We apply the second order accurate ``Koren'' reconstruction scheme as well as second order time-integration \citep{Porth2017}. 
To ensure the highest resolution on the inclined current sheet, we use a wedge-shaped static mesh refinement (SMR) whereby we enforce the highest refinement level within $\pm 45^\circ$ around the equatorial plane (see Figure \ref{fig:smrwedge} in Appendix \ref{app:numconv}).  
We use a base resolution with 128, 64, and 128 cells along the $\gls{sphradius}$, $\gls{sphpolar}$, and $\gls{sphazi}$ coordinates respectively and increase resolution by consecutively doubling the number of cells in each coordinate direction inside the SMR wedge.

To investigate numerical convergence (see Appendix \ref{app:numconv}) we performed 16 simulations (labeled SMR-SP) for the 4 different BH spins (indicated in the simulation label by the number after "SP") using 4 different increasingly larger effective resolutions (indicated in the simulation label by the number after "SMR").
In addition, we performed one simulation with five refinement levels (SMR5-SP99) with $\gls{spinpar} = 0.99$ i.e. an effective resolution of $2048\times 1024\times2048$ cells (in $\gls{sphradius}\times\gls{sphpolar}\times\gls{sphazi}$).
Finally, we performed two simulations (labeled SMR4-SP99-X0 and SMR4-SP99-X20) with $\gls{spinpar} = 0.99$ for $\chi_{0} = 0 ^{\circ}$ and $\chi_{0} = 20 ^{\circ}$ respectively to investigate the dependence on the initial inclination angle.

For numerical stability, we apply a floor model where we enforce a maximum plasma magnetization with identical radial dependence as in initialization except with $\gls{plmagn}_{0} = 100$ instead of $\gls{plmagn}_{0} = 50$.  
In addition, the Lorentz factor is capped at a maximum of $50$ to diminish the influence of unphysical cells that can sporadically appear throughout the evolution.
These can be caused by the lack of numerical accuracy when conserving momentum in the case of too low density leading to unphysically large Lorentz factors (see also the discussion how this is handled in contemporary codes in Section 4.1 of \cite{PorthChatterjeeEtAl2019a}).

%%%%%%%%%%%%%%%%%%%%%%%%%%%%%%%%%%%%%%%%%%%%%%%
%%%      Section Current sheet alignment    %%%
%%%%%%%%%%%%%%%%%%%%%%%%%%%%%%%%%%%%%%%%%%%%%%%
\section{Current sheet alignment}
% \section{Magnetospheric evolution and current sheet alignment}
\label{sec:csalignment} 
In this section we describe the evolution of the initially inclined split monopole current sheet. 
Figures \ref{fig:timesequencexz} and \ref{fig:timesequencexy} show a time-sequence for selected quantities at $\gls{time} = (45,111,270) \ \gls{gravradius} / \gls{spol}$ (all taken in the same phase of rotation) corresponding to the initial, intermediate, and late evolutionary phase of the highest resolution simulation SMR5-SP99. 
Figure \ref{fig:timesequencexz} displays 2D slices of the $\gls{cartx}\gls{cartz}$-plane ("side-on view" of the meridional plane) and Figure \ref{fig:timesequencexy} shows the $\gls{cartx}\gls{carty}$-plane ("top-down view" of the equatorial plane).  
\begin{figure*}[ht!]
\centering
\resizebox{\textwidth}{!}{ 
\plotone{figure_timesequence_xz.pdf} 
}
\caption{
A time-sequence of the highest resolution simulation (SMR5-SP99) with BH spin $\gls{spinpar} = 0.99$. 
We show square slices of the $\gls{cartx} \gls{cartz}$-plane ("side-on view") of 6 quantities. 
The rows show the co-moving plasma magnetization $\gls{plmagn}_{co}$, 
the plasma beta $\gls{plbeta}$,
the dimensionless temperature $\gls{dimtemp}$,
the Lorentz factor $\gls{lfacbulk}$, 
the radial magnetic field component times the square of the radial distance $\gls{Bcomp}^{\gls{sphradius}} \gls{sphradius}^{2}$ (to account for the scaling of the monopolar field), and
the radial 4-velocity $\gls{4velcomp}^{\gls{sphradius}}$. 
The 6 columns show for 3 different times $\gls{time} = (45,111,270) \ \gls{gravradius} / \gls{spol}$ (all taken in the same phase of
rotation) on the left a zoomed-out view and on the right a zoomed-in view.
The zoomed-out views cover a larger distance for later times as shown by the values on the axes at the bottom of the figures. 
In the panels the black circle is the BH, the white dashed line is the ergosphere, and the white dotted line is the inner light-surface. 
In the second row the white dashed dotted lines show the envelope of the event horizon current sheet inclination if it would radially propagate at the speed of light. 
In the fourth row the black arrowed lines represent the velocity field lines.
In the fifth row the black arrowed lines represent magnetic field lines. 
In the third column of the sixth row the black blocks represent the computational blocks showing the refinement wedge. 
}
\label{fig:timesequencexz}
\end{figure*} 
\begin{figure*}[ht!]
\centering
\resizebox{\textwidth}{!}{ 
\plotone{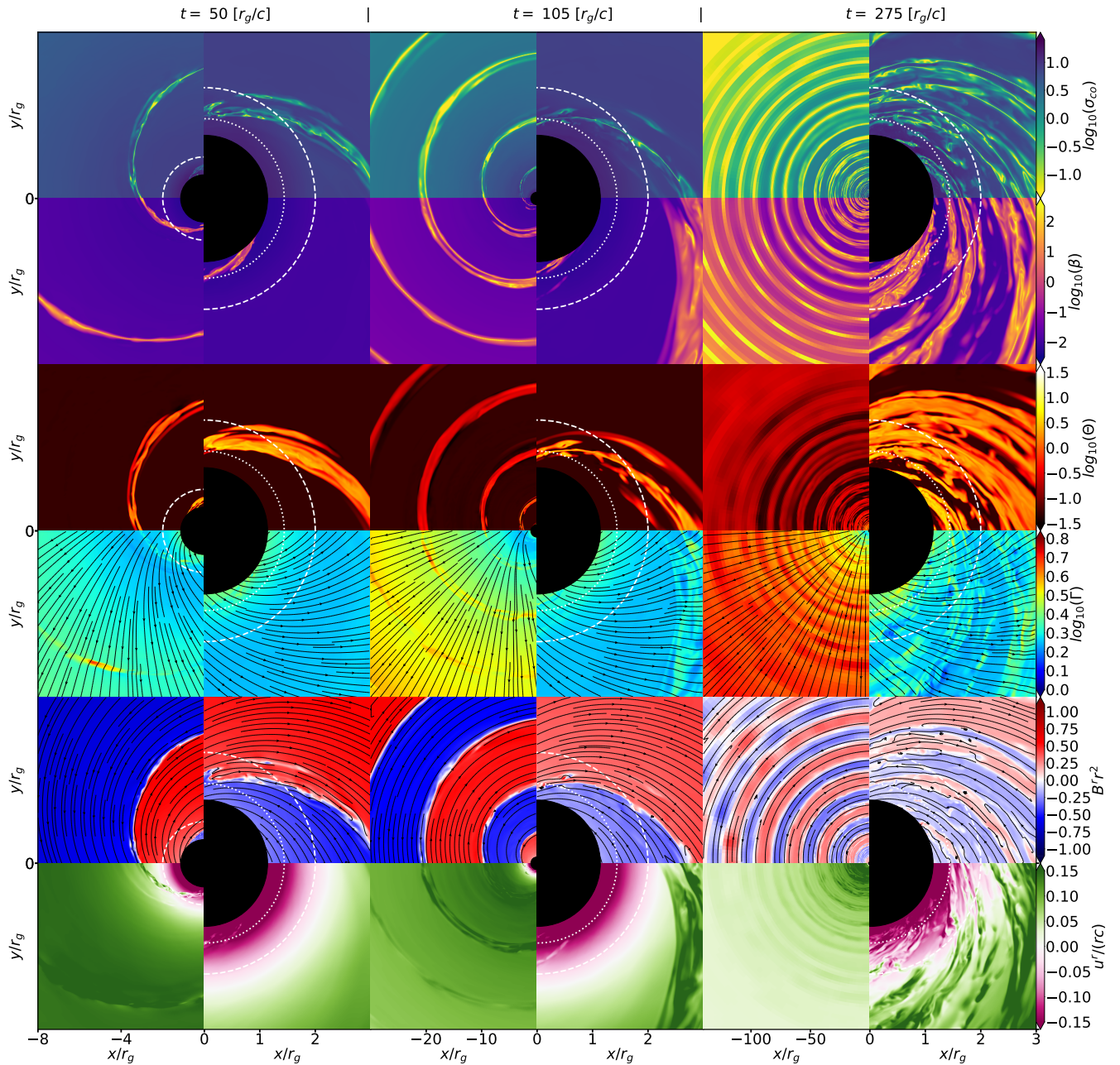} 
}
\caption{ 
A time-sequence identical to Figure \ref{fig:timesequencexz} of the $\gls{cartx} \gls{carty}$-plane (”top-down view”). 
Everything is the same as in Figure \ref{fig:timesequencexz} except that the panels in $\gls{carty}$-direction are off-centered from $\gls{carty}=0$ to better illustrate the spherical structure of the magnetosphere in the equatorial plane.
}
\label{fig:timesequencexy}
\end{figure*}

Due to spacetime rotation (Lense-Thrirring frame-dragging), the magnetic field lines start to co-rotate (i.e. rotation in the same direction as the spin of the BH), twist, and bend extracting energy from the BH spin from the ergosphere (white dashed lines) via the Blandford-Znajek (BZ) mechanism \citep{Blandford1977}. 
Contrary to the magnetic field lines being frozen into a neutron star surface, the magnetic field lines penetrating an event horizon are able to slide over it. 
The rotational velocity $\gls{angfreq}_{\textrm{F}}$ of the magnetic field, and therefore the current sheet, at the (outer) event horizon (black circles with outer event horizon radius $\gls{ehradius} = (1 + \sqrt{1 - \gls{spinpar}^{2} })\gls{gravradius} \approx 1.14 \gls{gravradius}$ for $\gls{spinpar}=0.99$) in our simulations is $\gls{angfreq}_{\textrm{F}} \simeq \gls{angfreq}_{\textrm{H}} / 2 $ with $\gls{angfreq}_{\textrm{H}} = \gls{spinpar} \gls{spol} / (2 \gls{ehradius})$ the BH angular velocity. 
This is in agreement with the prediction from the BZ solution \citep{Komissarov2004,Okamoto2006}.

The co-rotation of the inclined current sheet leads to an outward moving wave-like spiral structure (i.e. an Archimedean or Parker spiral) similar to the striped wind known from oblique pulsars \citep{Michel1971,Kirk2009}. 
In Figures \ref{fig:timesequencexz} and \ref{fig:timesequencexy} the striped wind current sheet is shown by the regions of high $\gls{plbeta}$ (in the second row) and low co-moving plasma magnetization $\gls{plmagn}_{\textrm{co}} =  b^{2} / (\gls{mdens} \gls{spol}^{2} ( 1 +  (\gls{polyidx}/(\gls{polyidx} -1))\gls{dimtemp}  )  ) $ (in the first row) with $b$ the co-moving magnetic field.
The striped wind structure occurs within an angular sector centered around the rotational equator on global scales comparable to the BH size (i.e. $\sim \gls{gravradius}$). 
The plasma magnetization, except in the current sheet, is $\gls{plmagn}_{\textrm{co}} \gtrsim 10$ such that for this magnetized wind the magnetic stresses accelerate plasma to a bulk Lorentz factor $\gls{lfacbulk} \sim \sqrt{\gls{plmagn}_{\textrm{co}}}$ (see fourth row in Figures \ref{fig:timesequencexz} and \ref{fig:timesequencexy}).
In addition, the dissipation of magnetic energy into plasma energy through reconnection  (effectively changing the adiabatic index)  creates a radial outward pressure gradient that may enhance the radial bulk acceleration.
On scales that are comparable to the current sheet thickness (i.e. smaller scales than the global scales of the striped wind) the current sheet is not smoothly-shaped but locally corrugated (see Figure \ref{fig:timesequencexz}). 
Inside the current sheet we find local magnetic X-lines (i.e. X-points) through which reconnection occurs leading to the formation of flux tubes (i.e. plasmoids) \citep{Bhattacharjee2009,Loureiro2012}.
The plasma magnetization in regions upstream from the current sheet make the plasma nearly force-free and results in an Alfv{\'e}n speed $\gls{alfvensp} = \sqrt{\gls{plmagn}_{\textrm{co}} / (\gls{plmagn}_{\textrm{co}} + 1)} \gls{spol} \gtrsim 0.95 \gls{spol}$ such that reconnection occurs in the relativistic regime. 
Consequently, the plasma inside the current sheet heats up locally to $\gls{dimtemp} \gtrsim 10 \sim \gls{plmagn}_{\textrm{co}}$ and becomes hotter than the surrounding plasma (see the third row in Figures \ref{fig:timesequencexz} and \ref{fig:timesequencexy}).
Inside the current sheet magnetic tension accelerates plasma in the direction of the reconnection outflows.

As the system evolves we observe that the co-rotating current sheet on the event horizon starts to tilt toward the BH equatorial plane until it has completely aligned. 
In other words, the current sheet inclination angle $\chi (\gls{time})$ (between the normal to the current sheet plane and the BH spin axis) decreases over time to zero (see by comparing the second, fourth, and sixth column of Figure \ref{fig:timesequencexz}). 
We note that we observe this current sheet alignment in all our simulations varying in BH spin and numerical resolution and in our test simulations with an initial magnetic dipole geometry.

As a result of alignment the striped wind is sourced at the event horizon by an increasingly smaller opening angle over time, as shown in Figures \ref{fig:timesequencexz}. 
Due to the outward motion of the striped wind this time dependence of the opening angle at the event horizon translates into a radial dependence of the opening angle leading to an amplitude-dampened striped wind. 
This reduction of the opening angle is illustrated by the envelope (white dash-dotted lines in the second row of Figure \ref{fig:timesequencexz}) created by event horizon current sheet if it would radially propagate at the speed of light. 
Comparing the striped wind to this envelope shows that it moves near-ballistically away from the BH with approximately the speed of light for all radii (see the first and third columns of Figures \ref{fig:timesequencexz} and \ref{fig:timesequencexy}). 
So not only has the envelope a curved shape (in contrast to the case of a pulsar where it is comprised by two diverging straight lines), inevitably this envelope over time will not be fixed in space but near-ballistically move outward.

The plasma motion is such that for $\gls{sphradius} \gtrsim 5 \gls{gravradius}$ it moves approximately radial (see the velocity field lines in the fourth row of Figures \ref{fig:timesequencexz} and \ref{fig:timesequencexy}).
Furthermore, a stagnation surface located at a few $\gls{gravradius}$ from the BH spin axis separates inward moving plasma from the outward moving plasma (see the last row of Figures \ref{fig:timesequencexz} and \ref{fig:timesequencexy}). 
This region which is being vacated from matter represents the starved magnetospheric region (i.e. gap) \citep{Levinson2018} in which pair formation may provide the outflow of matter. 
Moreover, magnetized plasma can only move inward inside the inner light-surface with radius $(1/4) \gls{metrcomp}_{\gls{sphazi}\gls{sphazi}} \gls{angfreq}_{\textrm{H}}^{2} + \gls{metrcomp}_{\gls{time}\gls{sphazi}} \gls{angfreq}_{\textrm{H}} + \gls{metrcomp}_{\gls{time}\gls{time}} $ (white dotted lines) with $\gls{metrcomp}_{\mu \nu }$ the metric components \citep{Gralla2014}.
In addition, in the outward moving plasma Poynting flux is converted into plasma kinetic energy through the Lorentz force acting on the matter thereby increasing the (bulk) Lorentz factor of the plasma for larger radii (see fourth row in  of Figures \ref{fig:timesequencexz} and \ref{fig:timesequencexy}).

\begin{figure}[ht!]
\centering
\resizebox{\columnwidth}{!}{ 
\plotone{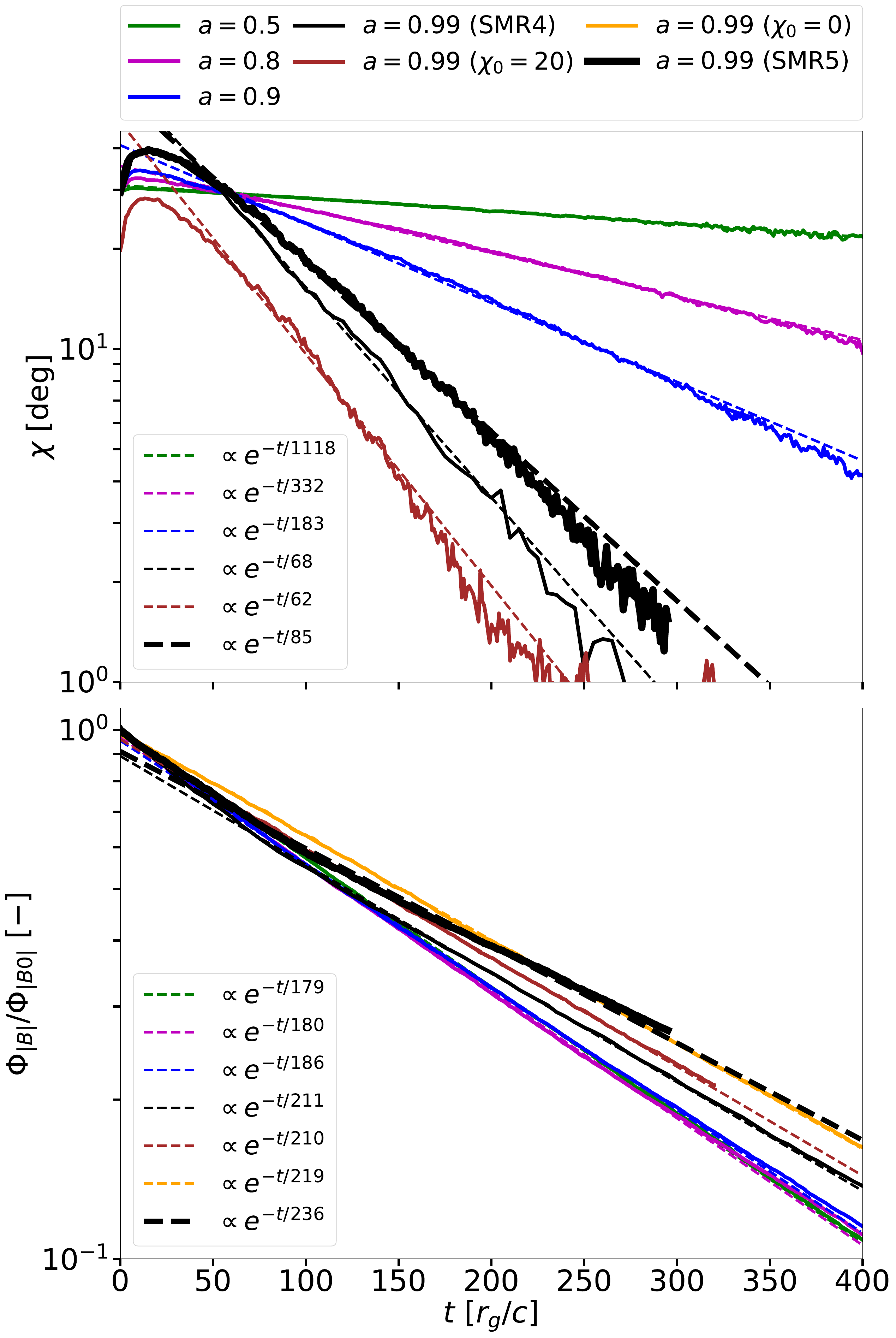}
}
\caption{
The time evolution of the current sheet inclination angle $\chi$ on the event horizon (top panel) and the total event horizon magnetic flux $\gls{magnfluxabs}$ of $|\gls{Bcomp}^{\gls{sphradius}}|$ normalized to its initial value (bottom panel) for BH spins $\gls{spinpar} = (0.5, 0.8, 0.9, 0.99)$. 
Next to the simulations with $\chi_{0} = 30 ^{\circ}$ for all spins we also included the simulations with $\chi_{0}= 20^{\circ}$ and $\chi_{0}=0^{\circ}$ with spin $\gls{spinpar}=0.99$ where the latter is not shown in the top panel as there is no alignment angle evolution.
Colored continuous lines are measured from the simulations and the matching colored dashed lines are exponential fits.
The top panel shows that the inclination angle decreases at an exponential rate, with timescale dependent on the BH spin where higher spin results in faster alignment.
The simulation with $\chi_{0}= 20^{\circ}$ aligns at an equal rate to the $\chi_{0}= 30^{\circ}$ case making the alignment timescale independent of $\chi_{0}$.
The bottom panel shows that for all spins and all $\chi_{0}$ the magnetic flux decays on a similar exponential timescale $\tau_{\Phi} \sim 2 \cdot 10^{2} \gls{gravradius} / \gls{spol}$.
}
\label{fig:alignmentandfluxvstime}
\end{figure}

In order to quantify the alignment we integrate the magnetic moment density of the current sheet over a spherical shell of single cell thickness localized on the event horizon while neglecting radial gradients. 
For this we take only the current density tangential to the event horizon (i.e. only current density components in $\gls{sphpolar}$- and $\gls{sphazi}$-direction).
The current sheet inclination angle $\chi(\gls{time})$ is then defined as the angle between this magnetic moment vector and the BH spin axis. 
This integrated measure works robustly down to $\chi\simeq 1^\circ$.
The top panel of Figure \ref{fig:alignmentandfluxvstime} shows the time evolution of $\chi(\gls{time})$ from our highest resolution simulations for each spin (continuous lines).
After the initial transient phase, which lasts for a few tens of $\gls{gravradius} / \gls{spol}$, for all spins the inclination angle decreases exponentially until the current sheet fully aligns (i.e. has an inclination angle of $\leq 1 ^{\circ}$).
The lowest considered spin $\gls{spinpar} = 0.5$ has an alignment timescale $\tau_{\chi}  \simeq 1123 \ \gls{gravradius} / \gls{spol}$ (green lines) which is more than an order of magnitude larger than that of the highest spin $\gls{spinpar} = 0.99$ making the alignment timescale evidently dependent on the BH spin, where higher spin results in faster alignment.
For the highest spin $\gls{spinpar} = 0.99$ (in the highest resolution simulation SMR5-SP99 indicated by the black thick continuous line) the current sheet aligns with an exponential alignment timescale $\tau_{\chi} \sim 88 \ \gls{gravradius} / \gls{spol}$ (indicated by the black dotted line) such that after a time evolution of $\simeq 360 \ \gls{gravradius} / \gls{spol}$, the alignment is complete ($\chi < 1 ^{\circ}$). 

\begin{figure}[ht!]
\centering
\resizebox{\columnwidth}{!}{ 
\plotone{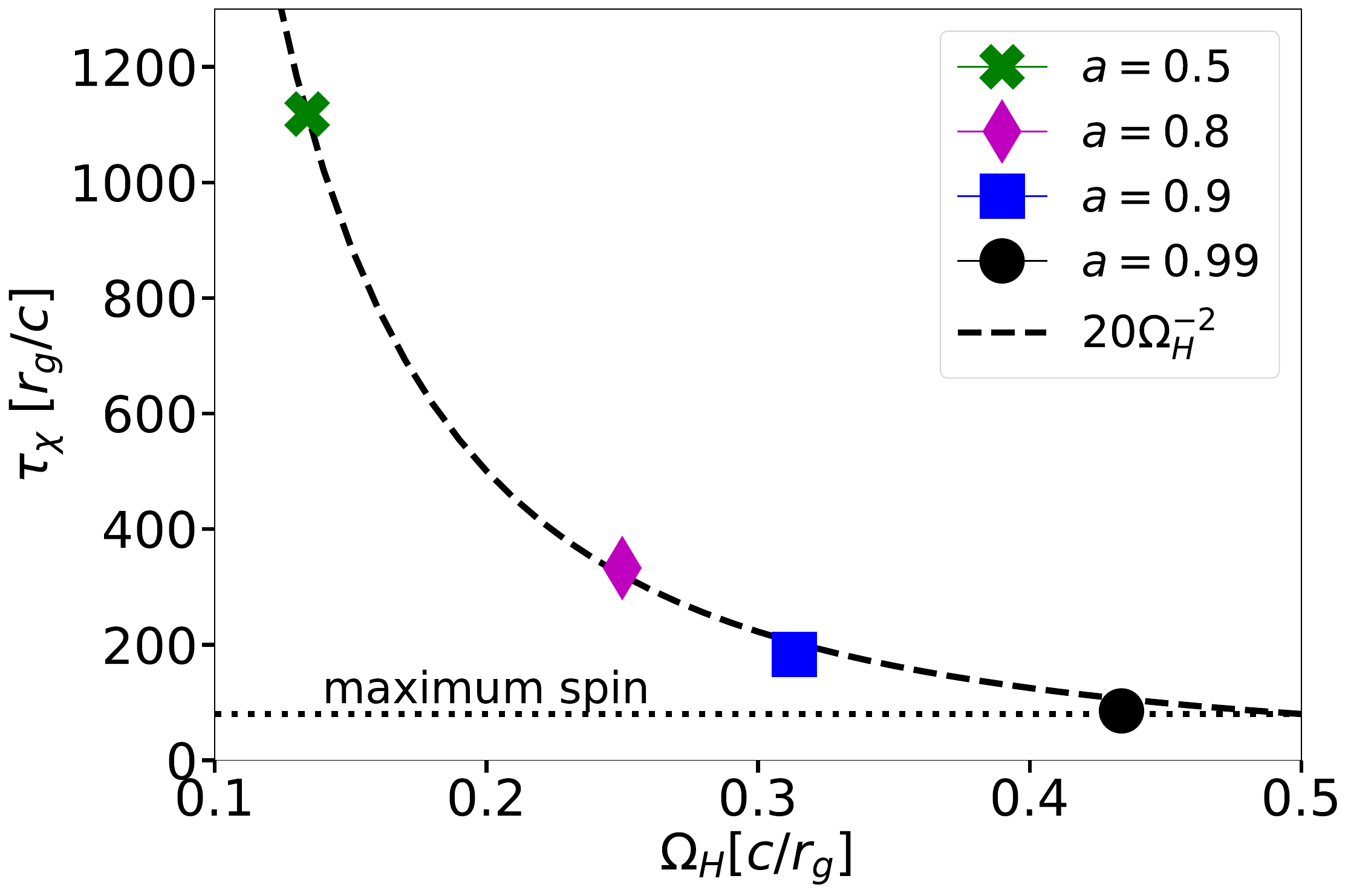}
}
\caption{
Exponential alignment timescales $\tau_{\chi}$ for all considered dimensionless BH spins $\gls{spinpar}=(0.5,0.8,0.9,0.99)$ as a function of the BH angular velocity $\gls{angfreq}_{\textrm{H}}$. 
The fit (dashed line) shows that the alignment timescale is inversely proportional to the square of the BH angular velocity.
}
\label{fig:tauvsomegah}
\end{figure}

To investigate the spin dependence we relate the alignment timescale to the BH angular velocity $\gls{angfreq}_{\textrm{H}}$ in units of $\gls{spol} / \gls{gravradius}$ shown in Figure  \ref{fig:tauvsomegah}.
By fitting the measured alignment timescales as a function of $\gls{angfreq}_{\textrm{H}}$ we find that $\tau_{\chi} \simeq 20 \gls{angfreq}_{\textrm{H}}^{-2}$ such that the alignment timescale is inversely proportional to the square of the BH angular velocity. 

For simulation SMR4-SP99-X20 with $\chi_{0} = 20 ^{\circ}$ (see the brown continuous line in Figure \ref{fig:alignmentandfluxvstime}) the exponential alignment timescale is similar to the matching case with $\chi_{0} = 30 ^{\circ}$ (see by comparing the brown dashed line to the thin black line). 
Therefore, we conclude that the alignment timescale is independent of the initial inclination angle and our results can be extrapolated to any inclination angle.

By comparing the electromagnetic output power of our simulations to the BZ power scaling \citet{Blandford1977,Tchekhovskoy2010a} while considering the instantaneous event horizon penetrating magnetic flux (i.e. taking flux decay into account, discussed in the next section), we conclude that the current sheet inclination angle and its alignment has no significant effect on the power extraction (see Appendix \ref{app:outputpower}). 
This is in contrast to the spindown power dependence on the inclination angle in an obliquely rotating NS \citep{Spitkovsky2006} in which the field lines are anchored into the NS surface.

%%%%%%%%%%%%%%%%%%%%%%%%%%%%%%%%%%%%%%%%%%%%%%%
%%%        Section Magnetic flux decay      %%%
%%%%%%%%%%%%%%%%%%%%%%%%%%%%%%%%%%%%%%%%%%%%%%%
\section{Magnetic flux evolution}
\label{sec:magnfluxdecay}
During an initial transient phase we observe that the event horizon magnetic field flux redistributes over several $\gls{gravradius} / \gls{spol}$ into a non-uniform configuration, in which the radial magnetic field component $\gls{Bcomp}^{\gls{sphradius}}$ peaks at the rotational poles, is lowest at the equator and is invariant in the azimuthal direction (see the second column of the fifth row in Figure \ref{fig:timesequencexz}). 
The inclined current sheet sits on top of this event horizon flux configuration as a magnetic field discontinuity. 
While \citet{Bransgrove2021} attributed this non-uniformity to a remnant of the initial dipole geometry, we observe that it also occurs in our split monopole simulations and therefore we conclude that it is related to the spacetime rotation. 
The degree of non-uniformity depends on the BH spin where higher spin leads to a higher degree of non-uniformity and zero spin to a completely uniform case.
We quantify the non-uniformity similarly to \citet{Bransgrove2021} with a non-uniformity factor $k$, where $k=1$ represents a uniform distribution and for $k>1$ more flux is concentrated at the poles than at the equator. We find that while for our lowest spin ($\gls{spinpar}=0.5$) $k \sim 1$ our highest spin ($\gls{spinpar}=0.99$) has $k \sim 1.2$.

Magnetic reconnection causes magnetic field lines (black arrowed lines in the fifth row of Figures \ref{fig:timesequencexz} and \ref{fig:timesequencexy}) from the upstream regions to flow toward the current sheet, thereby dissipating magnetic field energy.
Due to magnetic reconnection the magnetic flux on the event horizon decays and, therefore, also the radial magnetic field $\gls{Bcomp}^{\gls{sphradius}}$. 
In order to quantify the event horizon magnetic flux decay we consider a quantity which is closely related to the magnetic flux and which we define as 
$\gls{magnfluxabs} = \int_{0}^{2 \pi} \int_{0}^{\pi} |\gls{Bcomp}^{\gls{sphradius}}| \sqrt{\gls{spmetricdet}} d\gls{sphpolar} d\gls{sphazi}$ with $\gls{spmetricdet}$  the spatial metric determinant 
\footnote{
We define this quantity over the whole spherical surface, while in the literature this is sometimes defined over a single hemisphere.
}.
The bottom panel of Figure \ref{fig:alignmentandfluxvstime} shows the time evolution of $\gls{magnfluxabs}(\gls{time})$ normalized to its initial value $\Phi_{|\textrm{B0}|} =  \gls{magnfluxabs}(\gls{time}=0)$ for all simulated spins (continuous lines).
For all spins and initial inclination angles $\gls{magnfluxabs}$ decreases exponentially with a similar timescale $\tau_{\Phi} \sim 2 \cdot 10^{2} \gls{gravradius} / \gls{spol}$ ranging between $179 \gls{gravradius} / \gls{spol}$ and $236 \gls{gravradius} / \gls{spol}$. 
This means that after a time of  $\sim 460  \ \gls{gravradius} / \gls{spol}$ approximately $10 \%$  of the initial event horizon flux is left. 
The decay rate is exponential even after alignment has taken place and, in fact, is independent of the current sheet inclination angle.

We note that for higher spin (and higher resolution) the flux decay seems to be slightly slower. In order to investigate the differences between these flux decay rates more thoroughly, resistive (and even higher resolution) simulations are required.

%%%%%%%%%%%%%%%%%%%%%%%%%%%%%%%%%%%%%%%%%%%%%%%
%%%   Section Discussion and Conclusions    %%%
%%%%%%%%%%%%%%%%%%%%%%%%%%%%%%%%%%%%%%%%%%%%%%%
\section{Discussion and Conclusions} 
\label{sec:discandconc}

\subsection{The origin of current sheet alignment?}
The origin of alignment is still unclear. 
However, we now discuss briefly the physical aspects that may play a role in the alignment and offer our thoughts on several avenues that may lead to more insight.

First we point out some similarities and differences between our BH case and the case of a NS. 
For both vacuum and plasma-filled NS magnetospheres with dipolar magnetic fields it has been shown that the inclination angle between magnetic and rotational axes aligns over time due to the torques acting between the magnetosphere and stellar surface \citep{Deutsch1955,Goldreich1970,Michel1970,Philippov2014}. 
In contrast, it has been shown that for an obliquely rotating NS with a uniform split monopole magnetic field there is no temporal evolution of the inclination angle (i.e. no current sheet alignment) as there are no alignment torques \citep{Bogovalov1999}.
However, in the BH case the event horizon is not a perfectly conducting boundary, but magnetic field lines are allowed to move.
% allows magnetic field lines to slip. 
Therefore, the magnetic field is free to evolve which we see in the form of an event horizon magnetic flux redistribution and current sheet alignment.

The flux redistribution leads to a non-uniform configuration where the field strength peaks at the poles, is lowest at the equator and is invariant in the azimuthal direction (i.e. is axissymetric). 
This means that, in some sense, the magnetosphere is aligned with the BH spin axis except for the misaligned current sheet. 
The fact that we measure no significant effect of the current sheet inclination angle and its alignment on the power extraction through the BZ mechanism, indicates that an aligned magnetosphere and current sheet alignment can be thought of as two separate phenomena.
Individual sections of the inclined current sheet are located in regions with a magnetic field strength gradient in polar direction and sections at higher latitude are surrounded by a stronger magnetic field.
During the alignment the current sheet migrates towards the equator, yet the quasi-steady (i.e. except for magnetic dissipation) flux non-uniformity is roughly retained.
In other words, the alignment occurs in the form of a migrating magnetic field discontinuity (i.e. the current sheet) over a maintained axissymmetric non-uniform flux distribution.

This may suggests that the current sheet could be considered as a disk of matter in a differentially rotating spacetime that is being torqued into alignment.
This is reminiscent of Bardeen-Petterson alignment in accretion disks \citep{Bardeen1975}.  
The Bardeen-Petterson alignment mechanism depends on disk warping via differential spacetime rotation and outward transport of (misaligned-) angular momentum through viscosity. It is conceivable that the resolved instabilities in the current sheet facilitate the effective viscosity needed for this process.  

Finally, we note that in this work we have correlated the BH spin dependence of the alignment timescale to the spin dependent redistribution of the event horizon magnetic flux, where higher spin leads to both faster alignment and larger flux non-uniformity. 
This might suggest that this flux non-uniformity is ultimately related to alignment of the current sheet.

Although this discussion does not provide a physical explanation of the origin of alignment, it allows possible guidance for future research into this phenomenon.

\subsection{Observational signatures}

\paragraph{Precursor} 
The birth of a BH can lead to a number of transient electromagnetic and gravitational signatures. 
The perturbation of the magnetosphere may cause a monster shock \citep{Beloborodov2023} leading to emission preceding any observational signature created during the subsequent evolution of the magnetosphere (e.g., \citet{Nathanail2017}) considered in this work.
In the following, we discuss the possibility of a observational signature resulting from the current sheet alignment.

\paragraph{Striped wind characteristics}
Our case may produce observable emission similarly to a current sheet beyond the light cylinder of a pulsar. 
On top of the periodic rotation the alignment of the current sheet changes the inclination angle, leading to a striped wind with an progressively smaller opening angle.  
Moreover, event horizon magnetic flux loss leads to weakening of the magnetic field strength in the outward moving wind.

\begin{figure}[ht!]
\centering
\resizebox{\columnwidth}{!}{ 
\plotone{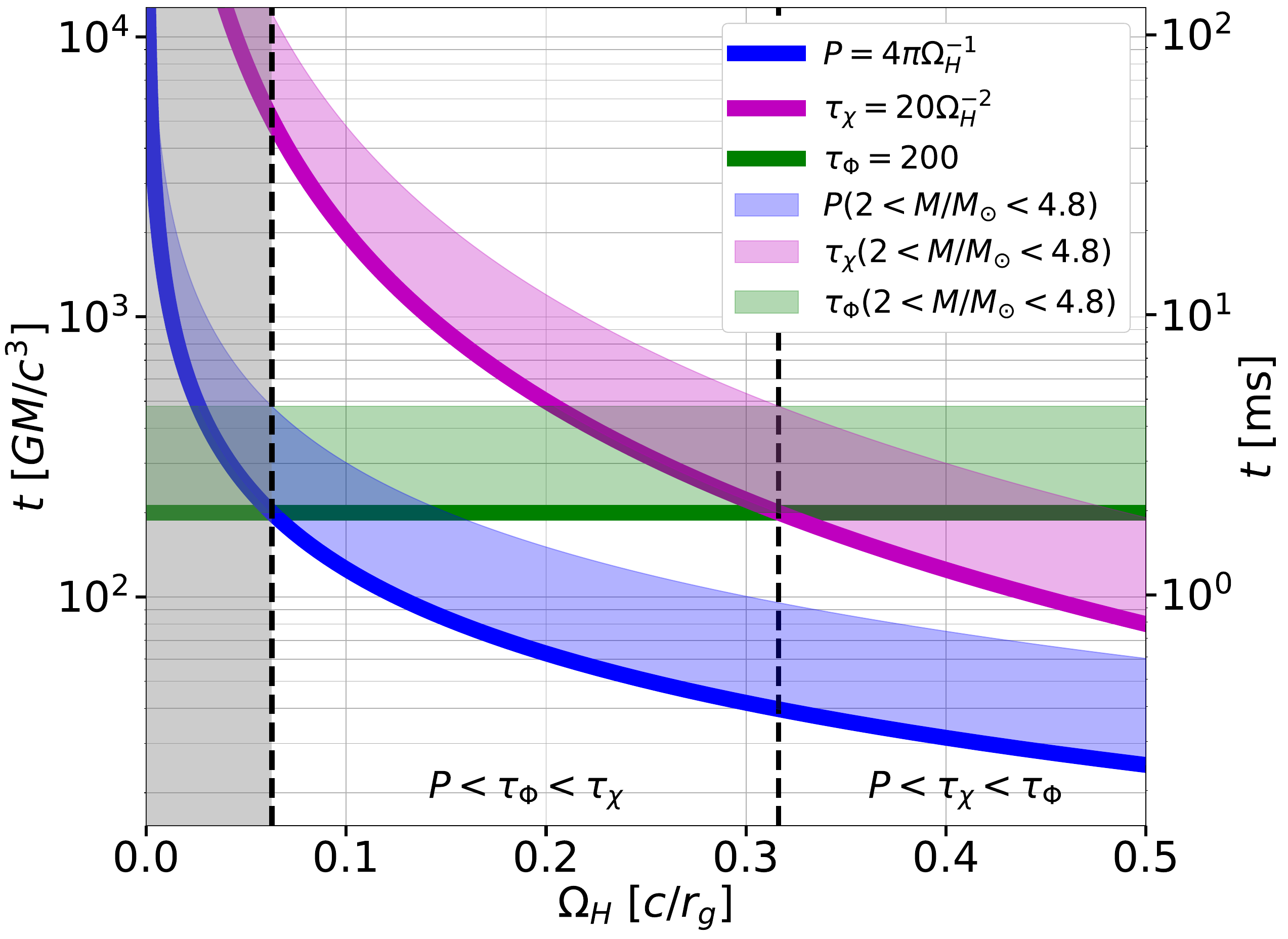}
}
\caption{The period $\gls{period}$, alignment timescale $\tau_{\chi}$ and decay timescale $\tau_{\Phi}$ as a function of BH angular velocity $\gls{angfreq}_{\textrm{H}}$. The colored thick lines are associated with the left vertical axis showing the dimensionless timescales. The colored shaded regions are associated with the right vertical axis showing the timescales in milliseconds restricted by the possible BH masses $2.0 \lesssim \gls{mass}    / \gls{mass}_{\odot} \lesssim 4.8$ formed from NS-NS mergers.
}
\label{fig:timesvsomega}
\end{figure}

This dampened striped wind is characterized by 3 timescales, the pulsation period $\gls{period}(\gls{angfreq}_{\textrm{H}}) \sim 2 \pi / (\gls{angfreq}_{\textrm{F}}/2) \sim  4 \pi / \gls{angfreq}_{\textrm{H}}$, 
the alignment timescale $\tau_{\chi}(\gls{angfreq}_{\textrm{H}})$ and 
the decay timescale $\tau_{\Phi}$ (all three timescales are shown in Figure \ref{fig:timesvsomega} by the thick colored lines associated with the left vertical axis) with $\gls{angfreq}_{\textrm{H}}$ in units of $\gls{spol} / \gls{gravradius}$. 
By inspection of the timescale expressions we can conclude that always  $\gls{period} < \tau_{\chi}$, meaning that the signal always will be periodic (before aligning). 
Additionally, we can conclude that the period and the alignment timescale are related such that, at a fixed BH mass a shorter period is always accompanied by faster alignment and vice versa.

We find 3 possible characteristic regimes of the signal.
For $\tau_{\Phi} \lesssim \gls{period} \lesssim \tau_{\chi}$ (which holds for $\gls{angfreq}_{\textrm{H}} \lesssim 0.06 $ and $\gls{spinpar} \lesssim 0.24$), assuming that our results can be extrapolated to smaller spins the signal will weaken due to magnetic flux loss before any of the other characteristics (rotation and alignment) are able to manifest themselves in the signal (shown by the shaded region in Figure \ref{fig:timesvsomega}).
For $\gls{period} \lesssim \tau_{\Phi} \lesssim \tau_{\chi}$ (which holds for $0.06 \lesssim \gls{angfreq}_{\textrm{H}} \lesssim 0.32$ and $0.24 \lesssim \gls{spinpar} \lesssim 0.91$) the signal is periodic, while weakening in strength faster than the alignment is able to significantly affect it. 
For $\gls{period} \lesssim \tau_{\chi} \lesssim \tau_{\Phi}$ (which holds for $\gls{angfreq}_{\textrm{H}} \gtrsim 0.32$ and $\gls{spinpar} \gtrsim 0.91$) the periodic variation in obliquity angle (at a given azimuth) has an envelope with an increasingly smaller opening angle while weakening only on longer timescales.

\paragraph{Black hole mass and spin from striped wind characteristics}
The expressions for $\gls{period}(\gls{angfreq}_{\textrm{H}})$ and $\tau_{\chi}(\gls{angfreq}_{\textrm{H}})$ allow us to determine expressions for the BH mass, 
\begin{eqnarray}
    \gls{mass} 
    \sim 
    26  
    \frac{(\gls{period} [\textrm{ms}])^{2}}{\tau_{\chi}[\textrm{ms}]}\gls{mass}_{\odot}    
\end{eqnarray}
with $\gls{period}$ and $\tau_{\chi}$ in milliseconds, and for the dimensionless BH spin parameter, 
\begin{eqnarray}
    \gls{spinpar}
    \sim
    \frac{20 \pi \tau_{\chi} \gls{period} }{( 100 \gls{period}^{2} +  \pi^{2} \tau_{\chi}^{2} )} 
\end{eqnarray} 
which are applicable to all regimes discussed above. 
If the period and alignment timescale can be measured, the spin and mass of the BH can directly be calculated using these expressions.

\paragraph{Timescale restrictions from black hole mass and spin bounds on NS-NS mergers}
Although limitations to the possible mass of a NS depend on the still unknown equation of state of its nuclear matter, various studies have been able to put bounds on it.
The mass of a remnant BH formed from a NS-NS merger is then, in principle, bounded on the lower end by a gravitationally collapsing maximum-mass non-rotating NS \citep{Rezzolla2018} and on the higher end by the (unlikely) merger of two supra-massive NSs \citep{Falcke2014} that are each maximally spinning, such that the range of masses we consider is $2.0 \lesssim \gls{mass}   / \gls{mass}_{\odot} \lesssim 4.8$.
The bounds on the BH mass puts limits on the (dimensional) range of the 3 characteristic timescales of our signal as shown in Figure \ref{fig:timesvsomega} by the shaded colored regions associated with the right vertical axis (in milliseconds).
For example, a maximally spinning BH ($\gls{angfreq}_{\textrm{H}} = 1/2$) with the lowest possible BH mass ($\gls{mass} = 2 \gls{mass}_{\odot} $) creates a signal with the shortest possible period $\gls{period} \sim 0.25  \textrm{ms}$ and fastest possible alignment rate $\tau_{\chi} \sim 0.8 \textrm{ms}$.

More likely though, the spin is more moderate resulting in a longer (and slower) pulsating signal.  
The maximum spin (the mass shedding limit) of a rigidly rotating NS is constrained to $\gls{spinpar} < 0.81$ setting a bound on the resulting BH spin \citep{AnnalaGordaEtAl2022}.  
Since there is plenty of orbital angular momentum available, some authors have argued that the NS-NS merger remnant should be near the mass-shedding limit \citep{PiroGiacomazzoEtAl2017}.
Moreover, it has been shown that a collapsing NS will lead to a BH with spin $\gls{spinpar} < 0.68$ \citep{Lo2011,Most2020}.
Thus assuming e.g. for a typical $\gls{spinpar} = 0.6$ (i.e. $\gls{angfreq}_{\textrm{H}} = 0.17$) and $\gls{mass} = 2 \gls{mass}_{\odot} $ the alignment timescale is $\tau_{\chi} \sim 7 \textrm{ms}$ with a period $\gls{period} \sim 0.74 \textrm{ms}$.

\paragraph{Characteristic timescales from timed pulses}
We can qualitatively describe the expected observational signature of the current sheet emission by using a simplified model. 
We assume (among other simplifications) that 
the direction of emission is along the plasma velocity direction due to relativistic beaming, 
that the plasma motion is radial for $\gls{sphradius} \gtrsim 5 \gls{gravradius}$ (see the fourth row in Figures \ref{fig:timesequencexz} and \ref{fig:timesequencexy}), 
that the emission originates from a fixed radius, e.g. from  $\gls{sphradius} = 5 \gls{gravradius}$ (being close to the BH as energy density drops quickly with radius), and 
that the emission can escape the magnetosphere toward an observer.
Then a (fixed) observer receives the emission in the form of timed pulses whenever the current sheet inclination angle at the emission radius matches that of the observer.

\begin{figure}[ht!]
\centering
\resizebox{\columnwidth}{!}{ 
\plotone{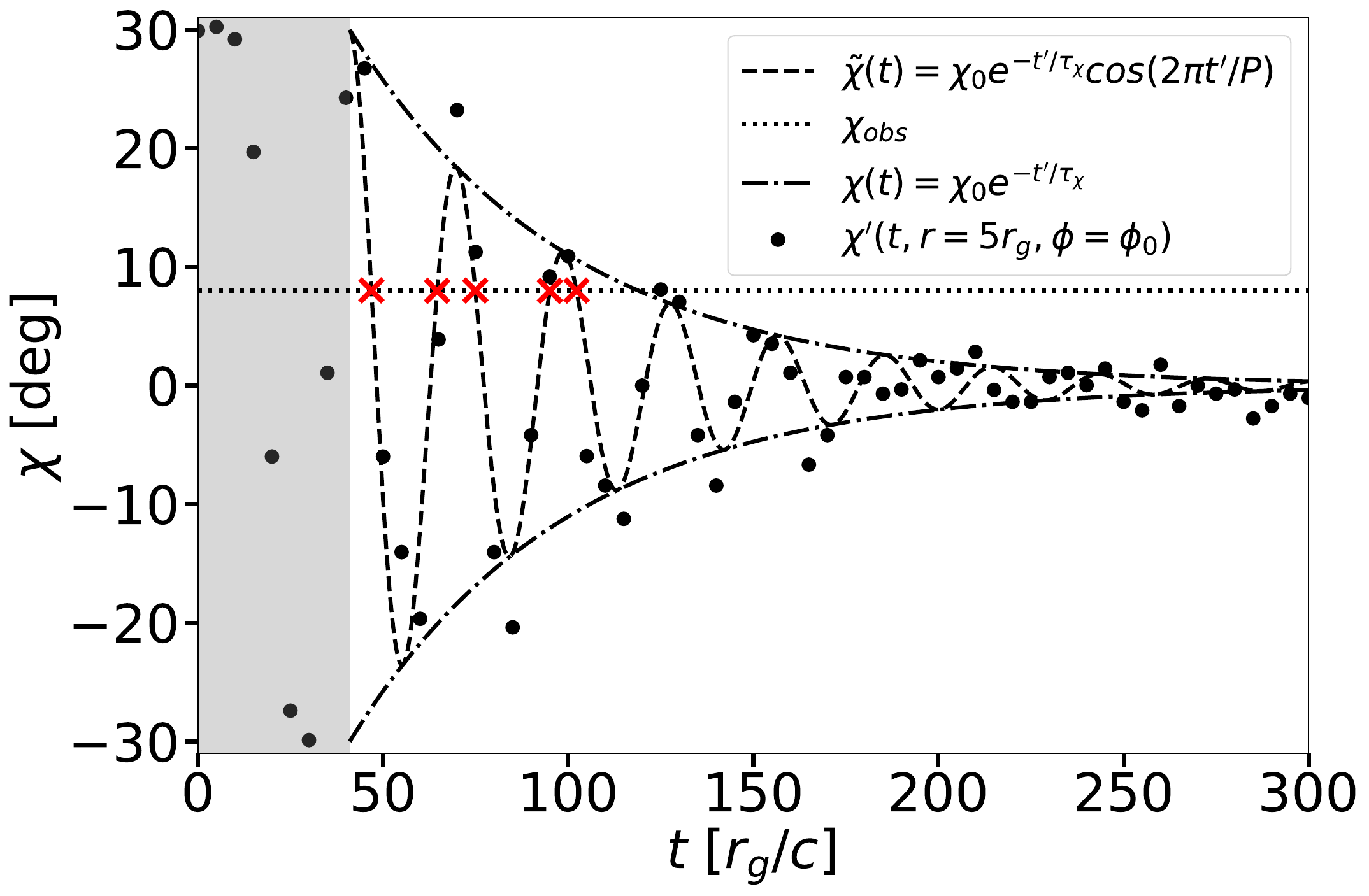}
}
\caption{An example of the general shape of an emitting striped wind current sheet described by its inclination angle $\tilde{\chi}$ (dashed lines) for a fixed azimuth with a shrinking envelope (dashed dotted line) described by the current sheet inclination angle $\chi$. 
A fixed observer at $\chi_{\textrm{obs}}$ (dotted line) may receive emission in the form of timed pulses (red crosses) if $\tilde{\chi}$ matches with $\chi_{\textrm{obs}}$. 
The current sheet inclination angle $\chi'$ (black dots) measured by the maximum current density along $\gls{sphpolar}$-direction at $\gls{sphradius}  = 5 \gls{gravradius}$ at a fixed azimuth $\gls{sphazi}_{0}$ (from simulation SMR5-SP99) show variability with respect to the general striped wind motion due to the local corrugated motion of current sheet and go slightly out of phase due to $\gls{angfreq}_{\textrm{F}}$ not being exactly $ \gls{angfreq}_{\textrm{H}} / 2 $.
}
\label{fig:signal}
\end{figure}

The general shape of the emitting current sheet inclination angle $\tilde{\chi}$ (i.e. the striped wind) at a fixed azimuth can be modelled by a dampening sinusoidal 
as $\tilde{\chi}(\gls{time}) = e^{- \gls{time}' / \tau_{\chi}} \chi_{0}  \textrm{cos}(2 \pi \gls{time}'  / \gls{period} )$ with $\gls{time}' = \gls{time} - \gls{time}_{0}$ the time offset accounting for
% or the arrival time at a particular radius and 
an initial transient phase for which this fitting formula is not valid.  
Here, the term $e^{- \gls{time}' / \tau_{\chi}} \chi_{0} $ is the current sheet inclination angle $\chi(\gls{time}) $.
In Figure \ref{fig:signal} (for conditions of simulation SMR5-SP99) $\tilde{\chi}(\gls{time})$ is shown by the dashed black line, the initial transient phase by the gray shaded region, and $\chi(\gls{time}) $ by the black dashed dotted line.
In addition, Figure \ref{fig:signal} shows the current sheet inclination angle $\chi'$ measured by the maximum current density along the $\gls{sphpolar}$-direction at $\gls{sphradius} = 5 \gls{gravradius}$ at a fixed azimuth $\gls{sphazi}_{0}$.
The variability of the measurements of $\chi'$ relative to modeled signal $\tilde{\chi}$ we attribute to the local corrugated motion of the current sheet causing deviations from the global striped wind motion. 
Moreover, the signal model $\tilde{\chi}$ assumes that $\gls{angfreq}_{\textrm{F}}$ is exactly $ \gls{angfreq}_{\textrm{H}} / 2 $, however the rotational velocity is actually slightly larger. The cumulative effect of this difference becomes apparent when at later times the measured signal goes slightly out of phase with the model.

At zeroth order, the separation between two consecutive odd or even pulses received by a fixed observer (shown by the dotted horizontal line) at an angle $\chi_{\textrm{obs}}$ gives the period $\gls{period}$ (e.g., from first to third red cross). 
However, as a result of alignment over time the measured time between two consecutive odd or even pulses shrinks, which is ultimately related to the alignment rate.
For the special case when assuming $\gls{period} \ll \tau_{\chi}$ it is possible to analytically determine $\gls{period}$ and $\tau_{\chi}$ (see Appendix \ref{app:analyticalpandtau}).

\paragraph{Emission and reconnection}
The emission is expected to be produced by radiation mechanisms that are similar to that of a pulsar in the current sheet beyond the light-cylinder. 
Merging plasmoids can eject fast magnetosonic waves which may escape as coherent electromagnetic radio emission \citep{Philippov2019}, strongly dependent on the (unknown) opacity..
Furthermore, reconnection-accelerated particles may lead to X-ray and gamma-ray emission \citep{Bai2010,Philippov2018b,Bransgrove2021}. 
In addition, we note that for the prediction of light curves caustic effects known from pulsars is likely to be relevant \citep{Bai2010}. 

We estimate the cooling regime of reconnection-accelerated particles by comparing the accelerating Lorentz force to the radiation back reaction force and assuming synchrotron cooling dominance.  
Given our restricted remnant BH mass inherited from NS-NS mergers results in $
\gls{Bcomp}_{\textrm{up}}^{3/2} \lesseqgtr 5 \gls{mplcty} 
$ 
for respectively the weak ($<1$), intermediate ($\sim 1$), or strong ($>1$) cooling regime with $\gls{Bcomp}_{\textrm{up}}$ the upstream magnetic field strength in units of Gauss and $\gls{mplcty}$ the pair multiplicity (i.e., the pair density in excess of Goldreich-Julian).
For a realistic scenario applicable to our work it is unlikely that $\gls{Bcomp}_{\textrm{up}}$ and $\gls{mplcty}$ are such that $ \gls{Bcomp}_{\textrm{up}}^{3/2} \lesssim 5 \gls{mplcty}$.
Therefore, we can conclude that high-energy emission is likely to be produced by particles in the strong cooling regime.

For both high-energy emission and radio emission to be observable they have to be able to to escape the magnetosphere.
Two effects that should be considered for this are high-energy photons that may annihilate into electron-positron pairs and radio emission that may be absorbed by the surrounding material. 
For a hypermassive NS born from an isolated progenitor (i.e. without an accretion disk or debris) the environmental absorption may likely be not an issue. 
In the case of a NS-NS merger it is highly likely that merger debris is present obscuring any emission. However, for long lived ($\sim$ day) hypermassive NS formed after a NS binary merger it can be argued that a large portion of the debris may have been accreted or has left the magnetosphere before the gravitational collapse to a BH \citep{Rowlinson2023}.
Moreover, it is suggested that post-merger debris mainly resides in the equatorial plane such that emission may escape at high latitudes \citep{Fujibayashi2018,Nathanail2020,Combi2023}. 
However, also for the long-lived hypermassive NS collapsing to a BH, there is a large uncertainty in the opacity due to pair production, depending both on the emitted photons and on the seed photons that may exist in the magnetosphere. We leave calculations of the emission escaping the magnetosphere for future work.

\subsection{Caveats}
We now comment on a few limitations of our work. 
Realistic BH magnetospheric plasma might be extremely magnetized ($\gls{plmagn}_{\textrm{co}} \gtrsim 10^{12}$)\citep{Philippov2021}, however current limitations of MHD simulations prevent us from reaching this regime (though see \citet{Phillips2023} for a potential remedy).
Therefore, we carried out simulations in the relativistic regime ensuring $\gls{plmagn}_{\textrm{co}} \gtrsim 10$, making the dynamics already very nearly force-free with $\gls{alfvensp} \gtrsim 0.95 \gls{spol}$. 
We then argue that our results are applicable to even more magnetized configurations.

Our simulations were performed in ideal MHD and rely on numerical resistivity for reconnection to occur. 
By carrying out a convergence study, we show that the dissipation and alignment timescales show signs of convergence for effective resolutions upward from $512 \times 256 \times 512$ cells in $\gls{sphradius} \times \gls{sphpolar} \times \gls{sphazi}$ directions (see Figure 5). 
% Only the lowest spin case ($\gls{spinpar}=0.5$), which has the longest alignment timescale still shows an increase of the alignment timescale at an effective resolution of  $1024 \times 512 \times 1024$ cells in $\gls{sphradius} \times \gls{sphpolar} \times \gls{sphazi}$ directions. 
This indicates that the alignment and dissipation in our study is not governed by the grid-scale physics.

In reality, the magnetized plasma is very likely collisionless, which implies that the reconnection rate is faster by a factor of $\sim 10$ compared to the collisional reconnection rate observed in MHD simulations \citep{Uzdensky2010,Comisso2016,Bransgrove2021}. 
Kinetic reconnection effects can in principle be modeled using a macroscopic resistivity in resistive GRMHD which mimicks the non-ideal effects in the reconnection region \cite{Selvi2023}.

The inherited magnetic dipole geometry transforming into the split monopole geometry used in this work could be a reasonable approximation for the collapse of a super- or hypermassive magnetized NS.  
However, for merger remnants gravitational effects likely will distort the magnetic field geometry before the BH is formed. 
In that case, merger simulations including the spacetime dynamics are necessary to provide more realistic magnetic and gravitational field dynamics.

\subsection{Main findings}
We show that an initially inclined split monopolar magnetospheric current sheet on the event horizon aligns with the equatorial plane of a spinning BH on an exponential timescale dependent on the BH spin.
We find that the alignment timescale is inversely proportional to the square of the BH angular velocity, where higher spin
results in faster alignment.
As a result of alignment the current sheet striped wind has an increasingly smaller opening angle while near-ballistically moving outward for all radii. 
Furthermore, we show that magnetic reconnection in the current sheet leads to event horizon magnetic flux decay on an exponential timescale $\tau_{\Phi} \simeq 2 \cdot 10^{2} \gls{gravradius} / \gls{spol}$ that is nearly the same for all considered BH spins $\gls{spinpar} = (0.5,0.8,0.9,0.99)$.  
We argue that the rotating, aligning, and magnetic flux reconnecting current sheet of an initially inclined split monopole black hole magnetosphere may lead to a unique and observable multimessenger electromagnetic counterpart to a gravitational wave event in the form of a pulsar-like black hole. 
% \textit{black hole pulsar}.

%%%%%%%%%%%%%%%%%%%%%%%%%%%%%%%%%%%%%%%%%%%%%%%
%%%       Section Acknowledgements          %%%
%%%%%%%%%%%%%%%%%%%%%%%%%%%%%%%%%%%%%%%%%%%%%%%
\section*{Acknowledgements} 
S.S. would like to thank Sasha Philippov for useful discussions and feedback.
B.R. would like to thank  Elias Most, Yuri Levin, Ashley Bransgrove, and Andrei Beloborodov for useful discussions. 
This publication is part of the project with No. 2021.001 which is financed by the Dutch Research Council (NWO).
B.R. is supported by the Natural Sciences \& Engineering Research Council of Canada (NSERC). 
L.S. acknowledges support from the DoE Early Career Award DE-SC0023015. This work was supported by a grant from the Simons Foundation (MP-SCMPS-00001470) to L.S. and B.R., and facilitated by Multimessenger Plasma Physics Center (MPPC).
The computational resources and services used in this work were partially subsidized by NWO domain science on the Dutch national e-infrastructure with the support of the SURF Cooperative (project No. NWO-2021.001) and partially provided by facilities supported by the Scientific Computing Core at the Flatiron Institute, a division of the Simons Foundation; and by the VSC (Flemish Supercomputer Center), funded by the Research Foundation Flanders (FWO) and the Flemish Government – department EWI.

%%%%%%%%%%%%%%%%%%%%%%%%%%%%%%%%%%%%%%%%%%%%%%%
%%%               References                %%%
%%%%%%%%%%%%%%%%%%%%%%%%%%%%%%%%%%%%%%%%%%%%%%%
\clearpage
\bibliography{references-jabref.bib, more-references.bib}{}
\bibliographystyle{aasjournal}

%%%%%%%%%%%%%%%%%%%%%%%%%%%%%%%%%%%%%%%%%%%%%%%
%%%                Appendix                 %%%
%%%%%%%%%%%%%%%%%%%%%%%%%%%%%%%%%%%%%%%%%%%%%%%
\appendix

%%%%%%%%%%%%%%%%%%%%%%%%%%%%%%%%%%%%%%%%%%%%%%%
%%%          Numerical convergence          %%%
%%%%%%%%%%%%%%%%%%%%%%%%%%%%%%%%%%%%%%%%%%%%%%%
\section{Numerical convergence}
\label{app:numconv}
\begin{figure}[ht!]
\centering
\resizebox{0.6\columnwidth}{!}{ 
\plotone{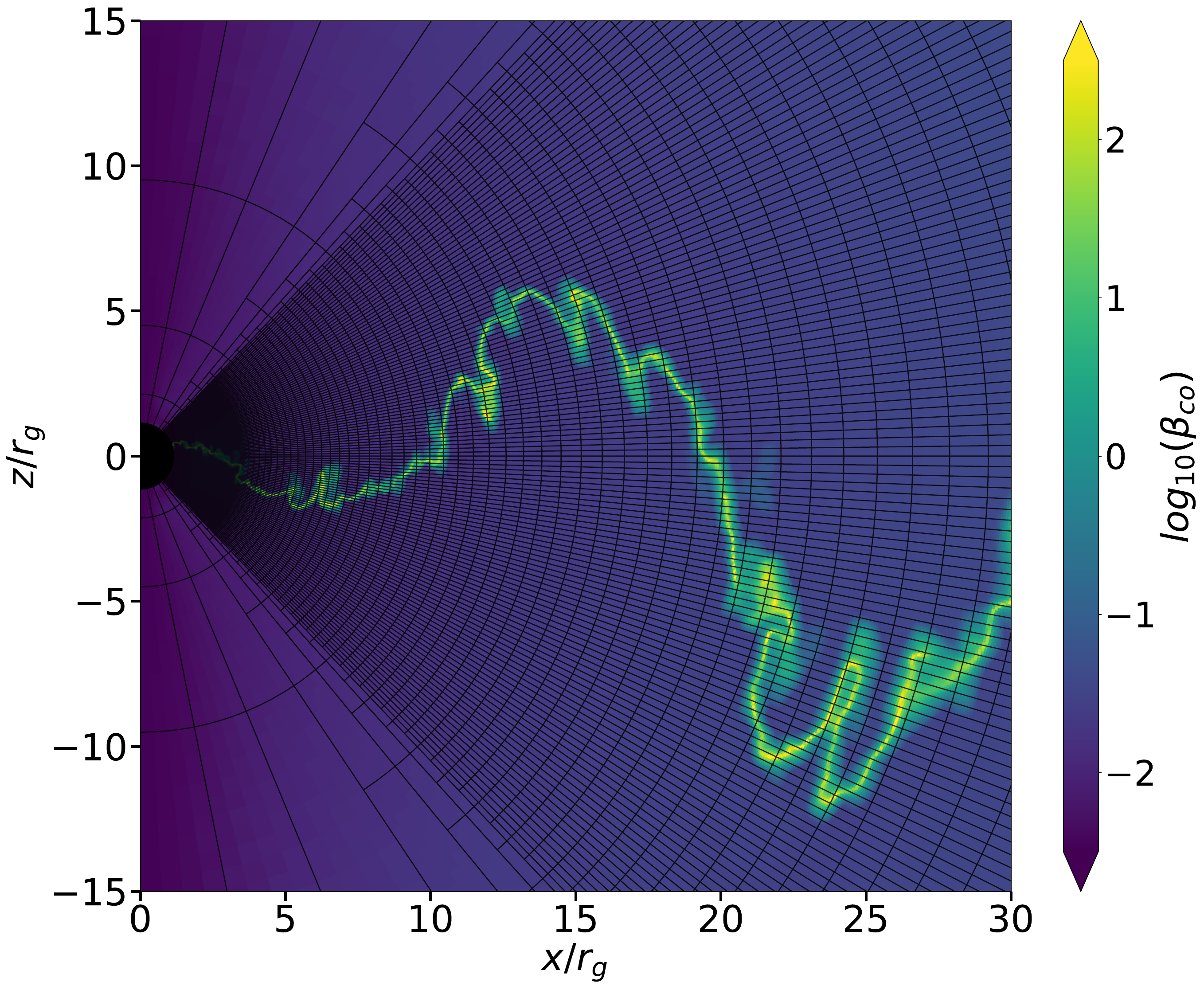}
}
\caption{A snapshot of the $\gls{cartx} \gls{cartz}$-plane in a "side-on-view" of the highest resolution simulation SMR5-SP99 showing a static mesh refinement structure (SMR). 
The wedge-shaped SMR structure has an half-opening angle of $45^\circ$ around the equatorial region and is displayed by gridlines representing the computational blocks. 
The computational blocks contain 20, 8, and 20 computational cells in respectively the $\gls{sphradius}$, $\gls{sphpolar}$, and $\gls{sphazi}$ coordinate direction.  
The colorscale shows plasma beta which inside the current sheet $\gls{plbeta}$ is higher than in the surrounding region. 
The current sheet at all times moves inside the SMR wedge, thereby ensuring that the highest resolution is imposed on the current sheet. 
}
\label{fig:smrwedge}
\end{figure}

To investigate the numerical convergence of our simulations we performed 16 simulations with initial inclination angle of $30^\circ$ for 4 different resolutions and 4 different black hole spins. In addition, we performed 1 high effective resolution simulation for $\gls{spinpar} = 0.99$ which we included in our convergence study.
We use static mesh refinement (SMR) to refine a wedge-like structure with half-opening angle of $45^\circ$ around the equatorial region to ensure the highest resolution in the region through which the current sheet moves (see Figure \ref{fig:smrwedge}).
\begin{figure}[ht!]
\centering
\resizebox{0.6\columnwidth}{!}{ 
\plotone{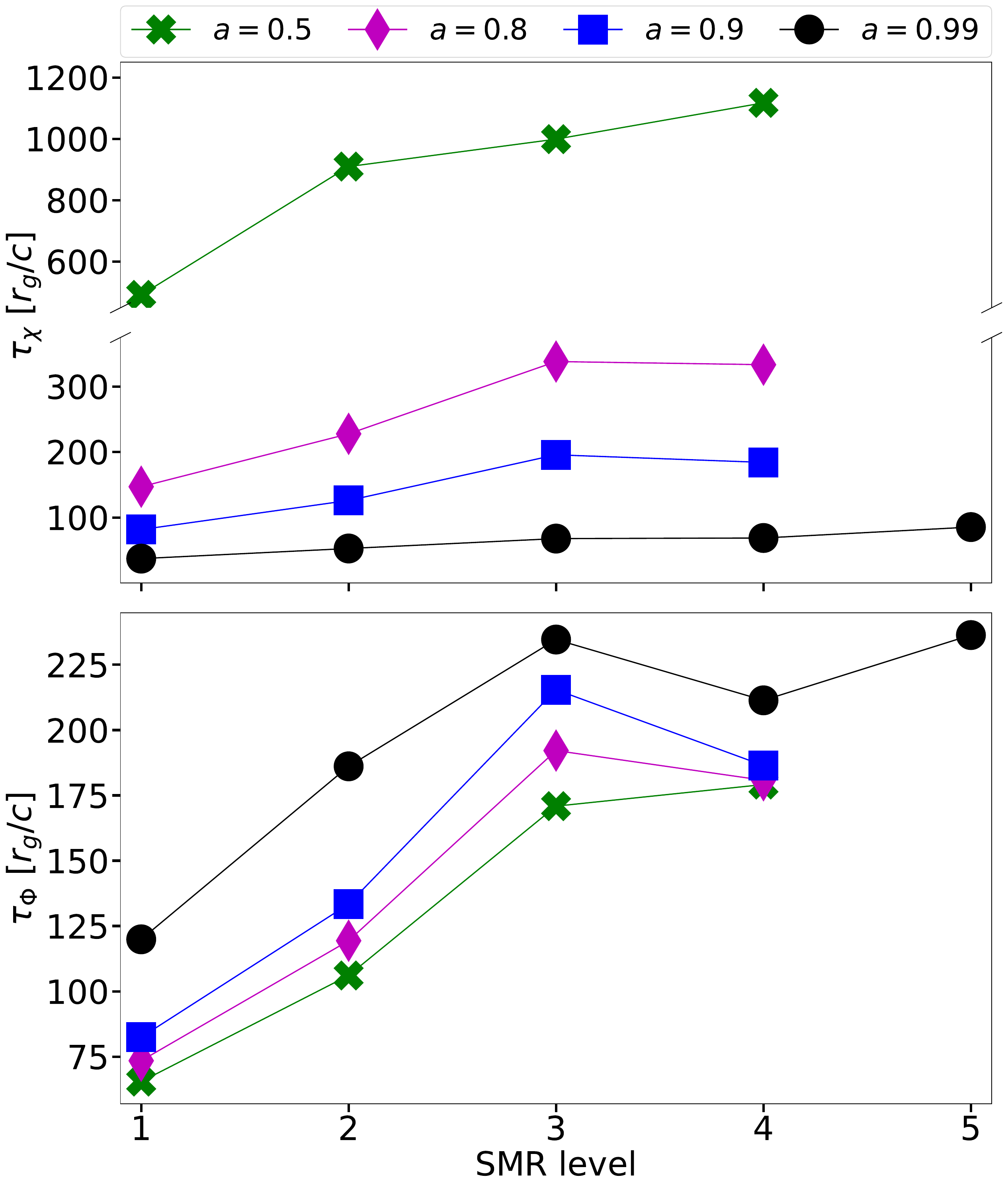}
}
\caption{
The current sheet alignment timescale $\tau_{\chi}$ and the event horizon magnetic flux decay timescale $\tau_{\Phi}$ as function of the numerical resolution indicated by the SMR level. This shows signs of numerical convergence beyond three SMR levels. 
}
\label{fig:convergence}
\end{figure} 
In our MKS coordinate system we use a base resolution with 128, 64, and 128 computational cells along the radial, polar, and azimuthal coordinates respectively.
To increase the resolution of our simulations we increase the level of SMR to double the amount of cells in each coordinate direction inside the SMR wedge. 
The two quantities we focus on are the characteristic current sheet alignment timescale $\tau_{\chi}$ of the current sheet inclination angle $\chi(\gls{time})$ and the characteristic decay timescale $\tau_{\Phi}$ of the magnetic flux on the event horizon $\gls{magnfluxabs}$ (see Section \ref{sec:magnfluxdecay} for a definition of this quantity).  

Figure \ref{fig:convergence} shows the characteristic timescales $\tau_{\chi}$ and $\tau_{\Phi}$ respectively in the top and bottom panel for all spins (indicated by the connected colored markers) as a function of the resolution (indicated by the SMR level).
The top panel shows that by increasing the resolution from the base level up to the third SMR level (effectively $256$ cells in polar direction) for all spins the alignment timescale increases. 
From the third SMR level up to higher resolutions (i.e. effective resolution higher than 256 cells in polar direction) we see the signs of convergence for the alignment timescale as for each spin the relative difference between the timescales of consecutive resolutions becomes smaller (i.e. the curves for each spin flatten out for higher resolutions). 
Furthermore, for all simulated resolutions the spin dependent hierarchy of the alignment timescale is preserved such that higher spin results in faster alignment.
Additionally, we note that the alignment timescale depends significantly on the black hole spin such that lower spin leads to larger timescales possibly requiring higher resolutions for convergence for the lower spin case.

The bottom panel shows that the event horizon magnetic flux decay timescale increases up to the third SMR level.
For resolutions up from the third SMR level (effectively $512\times 256\times512$ cells in $\gls{sphradius}\times\gls{sphpolar}\times\gls{sphazi}$ directions) we see a non-monotonous trend as the decay timescales of all spins of the fourth SMR level become lower (and the timescale in the single fifth level simulation is again slightly larger). 
In our simulations with resolutions higher than the third SMR level we observe that the current sheet is locally strongly corrugated and see the formation of X-points for reconnection and flux tubes (i.e. plasmoids). This may be caused by starting to resolve an instability at higher resolutions which is not resolved at lower resolutions. This may explain behavior of the decay timescale for resolutions upward from the third SMR level.  
Similarly to the alignment timescale, we see the decay timescale converging as for each spin the relative difference between the timescales of consecutive resolutions becomes smaller (i.e. the curves flatten). Furthermore, also here the spin dependent hierarchy is preserved such that higher spin results in slower decay. 
By comparing the highest resolution simulations for each spin we can also conclude that the flux decay timescales are similar and there is no significant spin dependence for all considered BH spins.
Although we can interpret the trend for the timescale upward from third SMR level as a sign of convergence, we cannot exclude that for resolutions higher than the fifth SMR level, which is computationally unfeasible for us at this point, the trend may change again. 
We note that there is a discrepancy of factor $\sim 2$ with the decay timescale presented by \citet{Bransgrove2021} for our initially aligned case. 
We have performed additional 2D simulations with an initial split monopole and dipole magnetic field geometry using resolutions comparable to our fourth SMR level, which show similar decay timescale to the 2D case of \citet{Bransgrove2021}. 
However, our simulations in 3D with varying initial tilt angle show a decay rate that is angle-independent, and in all cases a factor of two faster than the 3D result of \citet{Bransgrove2021} (for \citet{Bransgrove2021} 2D and 3D decay rates were the equal).
A more detailed investigation of this discrepancy and the differences between our approach using 3D ideal MHD in a logarithmic Kerr-Schild (KS) coordinate grid versus \citet{Bransgrove2021} using 3D resistive MHD in an equatorially squeezed (i.e., resulting in a significantly higher resolution for an equatorial current sheet) logarithmic KS coordinate grid are beyond the scope of this work. 

% Based on this numerical convergence study of our simulations we can conclude that our simulations show strong signs of convergence, in particular for the main conclusions of this work on the alignment of the magnetospheres.

%%%%%%%%%%%%%%%%%%%%%%%%%%%%%%%%%%%%%%%%%%%%%%%
%%%            BZ power scaling             %%%
%%%%%%%%%%%%%%%%%%%%%%%%%%%%%%%%%%%%%%%%%%%%%%%
\section{Comparison of energy flux to Blandford-Znajek power scaling}
\label{app:outputpower}
\citet{Blandford1977} showed that a magnetic field can extract black hole spin energy in the form of outgoing energy flux according to the Blandford-Znajek power scaling $\gls{power}_{\textrm{BZ}}$. 
Several expressions for $\gls{power}_{\textrm{BZ}}$ are listed in \citet{Tchekhovskoy2010a} which are valid for different ranges of black hole spin and are validated by numerical simulations.
The most general expression is an expansion in $\gls{angfreq}_{\textrm{H}}$ up to the sixth order, \citep{Tchekhovskoy2010a},
\begin{eqnarray}
    \gls{power}_{\textrm{BZ6}} (\gls{time})
    =
    \kappa \gls{magnfluxabs}(\gls{time})^{2} (\gls{angfreq}_{\textrm{H}}^{2} + \alpha \gls{angfreq}_{\textrm{H}}^{4} + \beta \gls{angfreq}_{\textrm{H}}^{6} )
\end{eqnarray}
with 
$\kappa$ a constant depending on the magnetic field geometry which is $1/(6 \pi)$ for our monopolar field,
$\gls{magnfluxabs}(\gls{time})$ our instantaneous magnetic flux,
$\alpha \approx 1.38$, and $\beta \approx -9.2$. Since $\gls{power}_{\textrm{BZ6}} \propto \gls{magnfluxabs}^{2}$ we use the instantaneous $\gls{magnfluxabs}(\gls{time})$ as the magnetic flux decays in our simulations.

\begin{figure}[ht!]
\centering
\resizebox{0.6\columnwidth}{!}{ 
\plotone{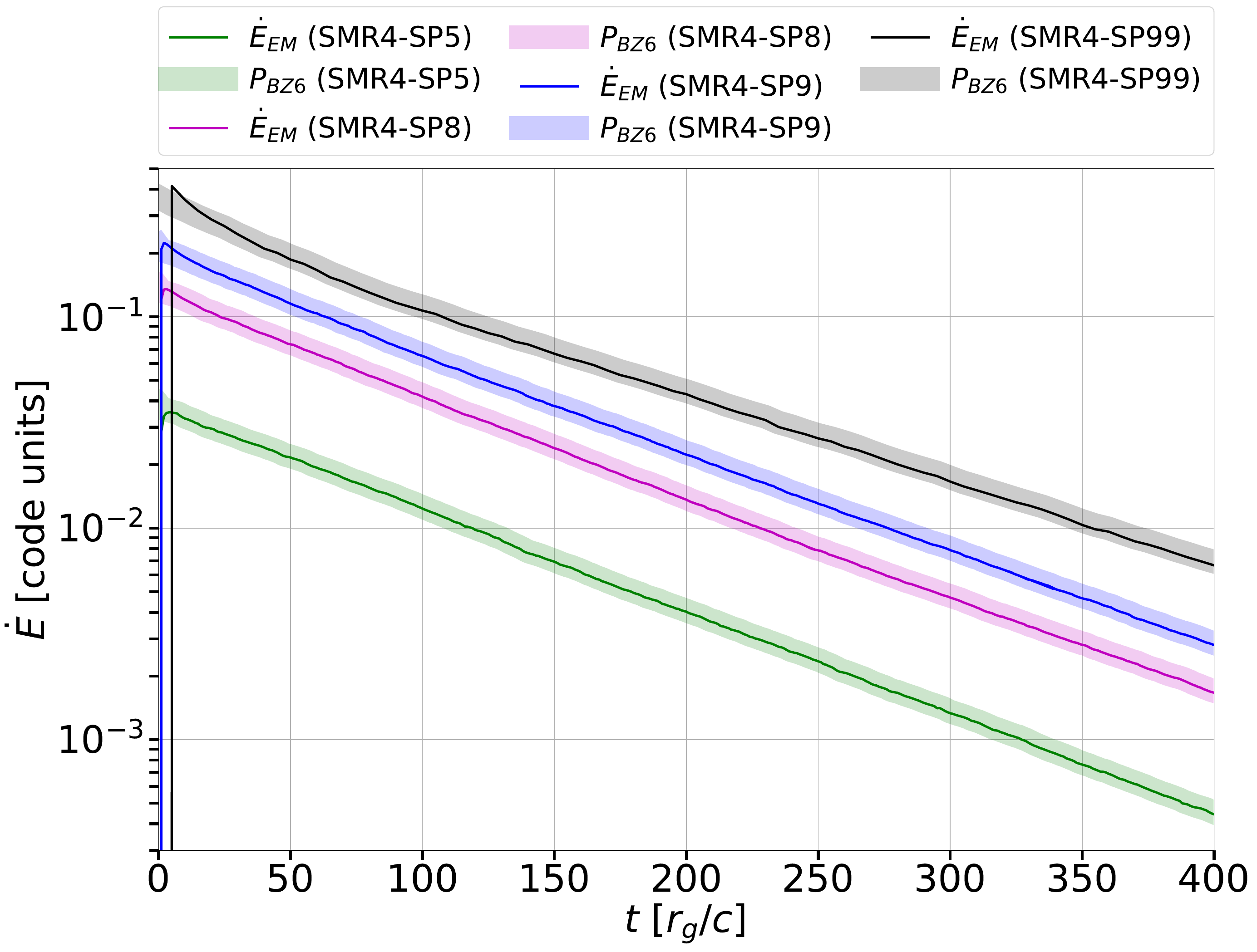}
}
\caption{
Time evolution of the measured electromagnetic radial energy flux $\gls{enrgyflux}_{\textrm{EM}}$ over the event horizon and the  corresponding instantaneous Blandford-Znajek power scaling $\gls{power}_{\textrm{BZ6}}$ for all 4 simulated spins $\gls{spinpar} = (0.5,0.8,0.9,0.99)$. For all spins during the entire evolution $\gls{enrgyflux}_{\textrm{EM}}$ is in excellent agreement with $\gls{power}_{\textrm{BZ6}}$ and there is no alignment dependence on the output power.
}
\label{fig:outputpower}
\end{figure}

In addition, in our simulations we measure the output power by measuring the electromagnetic radial energy flux $\gls{enrgyflux}_{\textrm{EM}}$ through a spherical shell at the event horizon (e.g., \citep{PorthChatterjeeEtAl2019a}),
\begin{eqnarray}
    \gls{enrgyflux}_{\textrm{EM}}
    =
    - \int_{\gls{sphazi} = 0 }^{\gls{sphazi} = 2 \pi }
    \int_{\gls{sphpolar} = 0 }^{\gls{sphpolar} =  \pi }
    (\gls{enrgmomtencomp}_{\textrm{EM}})^{\gls{sphradius}}_{\gls{time}} \sqrt{-\gls{metrtendet}}
    d\gls{sphpolar} d\gls{sphazi}
\end{eqnarray}
Here, $(\gls{enrgmomtencomp}_{\textrm{EM}})^{\mu}_{\nu} $ is $\mu \nu$-component of the electromagnetic part of the energy-momentum tensor and $\gls{metrtendet}$ the determinant of the metric. 
In Figure \ref{fig:outputpower} we show for all 4 simulated spins (for highest resolution simulations SMR4-SP5, SMR4-SP8, SMR4-SP9, SMR4-SP99) the time evolution of the measured $\gls{enrgyflux}_{\textrm{EM}}$ and the corresponding $\gls{power}_{\textrm{BZ6}}$.
This figure shows that for all spins during the entire evolution $\gls{enrgyflux}_{\textrm{EM}}$ and $\gls{power}_{\textrm{BZ6}}$ decrease exponentially. 
Furthermore, we verified that the difference between electromagnetic and total radial energy flux is negligible for all spins during the entire evolution implying a well magnetized system. 
Moreover, for all spins during the entire evolution $\gls{enrgyflux}_{\textrm{EM}}$ is in excellent agreement with $\gls{power}_{\textrm{BZ6}}$. 
Therefore, the current sheet inclination angle has no (significant) effect on the radial electromagnetic output flux and we conclude that the output power of the BH magnetosphere is not dependent on the alignment process.

%%%%%%%%%%%%%%%%%%%%%%%%%%%%%%%%%%%%%%%%%%%%%%%
%%%    Analytical solution of P and tau     %%%
%%%%%%%%%%%%%%%%%%%%%%%%%%%%%%%%%%%%%%%%%%%%%%%
\section{Analytical solution of the period and the alignment timescale for $P \ll \tau_{\chi}$}
\label{app:analyticalpandtau}
Here we determine the analytical solution for the special case of $\gls{period} \ll \tau_{\chi}$ from the pulse times $\gls{time}_{n}$ at the intersections between the model of the general shape of the striped wind and the observer (i.e. $\chi_{\textrm{obs}} = \tilde{\chi}(\gls{time})$). 
In this regime, the period may be approximated by bi-sequential pulses (i.e. between two consecutive odd or even pulse) such that $\gls{period} \approx \gls{time}_{n} - \gls{time}_{n+2}$. 
For a fixed period of the striped wind signal the time between pulses does not only depend on the fixed observer angle but also also changes due to alignment. 
At times $\gls{time}_{\textrm{I}} = (\gls{time}_{n+1}  - \gls{time}_{n}  )/ 2 $ and $\gls{time}_{\textrm{II}}=  (\gls{time}_{m+1}  - \gls{time}_{m}  )/ 2$ the observer inclination angles relative to the current sheet inclination angles are respectively 
$
( \chi_{\textrm{obs}} / \chi  )_{\textrm{I}} 
= 
\textrm{cos}( 2 \pi \gls{time}_{\textrm{I}} / \gls{period} )
$
and 
$
( \chi_{\textrm{obs}} / \chi  )_{\textrm{II}} 
= 
\textrm{cos}( 2 \pi \gls{time}_{\textrm{II}} / \gls{period} )
$.
Then this results in an alignment timescale 
$
\tau_{\chi}
=
(\gls{time}_{\textrm{II}} - \gls{time}_{\textrm{II}}) 
/
\textrm{ln} (  ( \chi_{\textrm{obs}} / \tilde{\chi}  )_{\textrm{II}}  / ( \chi_{\textrm{obs}} / \tilde{\chi}  )_{\textrm{I}}   )
$. 
As such, a minimal of 3 pulses are necessary to observationally obtain $\gls{period}$ and $\tau_{\chi}$.

\end{document}